\documentstyle[psfig]{mn}

\title[Intrinsic Properties of Galaxy Clusters in Different Cosmological 
Scenarios]{Properties of galaxy clusters in different cosmological 
scenarios. I. Intrinsic properties.}
\author[P.A.M. de Theije, E. van Kampen \& R.G. Slijkhuis]
 {Pascal A.M. de Theije$^{1}$\thanks{Present address: TNO Netherlands 
  Organization for Applied Scientific Research, The Hague, The Netherlands}, 
  Eelco van Kampen$^{2}$\thanks{Present address: Theoretical Astrophysics 
  Center, Juliane Maries Vej 30, DK--2100 Copenhagen, Denmark}, 
  Remco G. Slijkhuis$^{1}$ \\
  $^{1}$Sterrewacht Leiden, P.O.Box 9513, 2300 RA Leiden, The
        Netherlands \\
  $^{2}$Royal Observatory Edinburgh, Blackford Hill, Edinburgh EH9 3HJ, 
        Scotland}
\date{Accepted ; Received ; in original form }
\pagerange{\pageref{firstpage}--\pageref{lastpage}}

\begin{document}

\label{firstpage}

\maketitle

\begin{abstract}
We study the influence of the various parameters of scenarios of large--scale 
structure formation on properties of galaxy clusters, and investigate which 
cluster properties are most sensitive to these parameters. 
We present a set of large N--body simulations and derive the intrinsic 
properties of galaxy clusters in these simulations, which represent a volume 
of 256$^{3}$ $h^{-3}$ Mpc$^{3}$. 
The cosmological scenarios studied differ in either the shape of the power 
spectrum of initial fluctuations, its normalization, the density parameter 
$\Omega_{0}$, or the Hubble parameter $H_{0}$. 
Between each of the simulations, only one parameter is set differently, so
that we can study the influence of that parameter on the cluster properties.
The cluster properties that are studied are the mass, line--of--sight velocity 
dispersion, peculiar velocity, intrinsic shape, and orientation with respect 
to its surroundings. 

\par The present--day {\it r.m.s.} mass fluctuation on scales of $8 h^{-1}$ 
Mpc, $\sigma_{8}$, which is largely determined by the normalization of the 
initial power spectrum, has a large impact on the cluster properties. 
The latter, viz. the cluster number density, mass, line--of--sight velocity 
dispersion and peculiar velocity, are also determined by $\Omega_{0}$, though 
somewhat less. 
Other parameters, such as $H_{0}$, the tilt of the initial fluctuation 
spectrum, and the exact shape of this spectrum, are generally less important. 

\par Unlike the other cluster properties studied, the peculiar velocity is 
found to depend on all parameters of the formation scenario. 

\par In a companion paper the properties of the model clusters are compared to 
observations to try and discriminate between different cosmological scenarios. 
Using scaling relations between the average properties of the cluster sample 
and the 
parameters of the formation scenario, one may try and interpolate between 
the scenarios studied here in order to find the parameters of the scenario 
that is most consistent with the data. 
\end{abstract}



\section{INTRODUCTION}
\label{introduction}

Clusters of galaxies are the largest objects in the Universe that have 
recently collapsed, which makes them readily identifiable objects whose 
statistical properties can help to constrain scenarios for large--scale 
structure formation in the Universe. 

\par N--body simulations of large patches of the Universe were used 
extensively to compare the properties of the simulated clusters with 
observations. 
Frenk {\it et al.} (1990) calculated the distribution of velocity dispersions 
of clusters, identified both in 3--D and in 2--D. 
Jing \& Fang (1994) determined the cumulative distributions of mass, 
velocity dispersion, temperature of clusters as well as their space density 
in three scenarios, viz. standard CDM, low--$\Omega_{0}$ CDM, and a hybrid of 
CDM and HDM. 
Jing \& B\"orner (1995) determined the velocity dispersion profiles of 
clusters for seven cosmological scenarios. 
They found that the velocity dispersion profiles depend both on $\Omega_{0}$ 
and on the cosmological constant $\Lambda$. 
The profiles are steeper in a low--$\Omega_{0}$ scenario than for a high value 
of $\Omega_{0}$. 
This effect is weaker for a larger value of $\Lambda$. 
Mohr {\it et al.} (1995) carried out numerical simulations for three 
cosmological scenarios ($\Omega_{0}=1.0$, $\Omega_{0}=0.2$, 
$\Omega_{0}=0.2$ with $\lambda_{0}=\Lambda/3H_{0}^{2}=0.8$) with an effective 
power spectrum $P(k) \propto k^{-1}$ on cluster scales. 
They compared the projected X--ray shapes of the clusters in these scenarios 
to the observed shapes. 
They favoured the $\Omega_{0}=1$ scenario, although the authors admitted that 
some discrepancies remain. 
De Theije {\it et al.} (1995) used the galaxy distribution of 99 clusters to 
determine the projected ellipticities, and compared these to the results of 
N--body simulations of van Kampen (1994). 
They found that in the $\Omega_{0}=1$ CDM scenario clusters are too elongated 
with respect to real observed clusters. 
From a limited number of simulations for an $\Omega_{0} = 0.2$ CDM model, they 
concluded that clusters in this scenario are generally more nearly spherical 
than for $\Omega_{0} = 1.0$. 

\par N--body simulations were also used to study cluster alignments. 
Dekel, West \& Aarseth (1984) and West, Dekel \& Oemler (1989) concluded that 
the orientations of clusters of galaxies with respect to their neighbours 
provide a sensitive test for the formation of large--scale structure in the 
Universe. 

\par Many observational properties of clusters were studied in the recent past 
and much numerical work has been done. 
Most of the latter, however, concentrated on just one or two cluster 
properties. 
Also, in most studies up till now clusters were identified in the simulations 
in three dimensions, or background galaxies were not removed from the 
observations. 

\par In order to improve upon this state of affairs, we have run a set of 
cosmological N--body simulations with the following goals: 
(1) Evaluate the intrinsic properties of the clusters, i.e., of the groups of 
particles defined in 3--D that fulfill certain constraints. 
(2) Investigate the influence of the cosmological parameters on the cluster 
properties and study which cluster properties are mostly affected by varying 
the cosmology. 
(3) Obtain scaling relations between the average properties of the cluster 
sample and the parameters of the formation scenario to allow interpolation 
between the studied scenarios. 
(4) Compare the cluster properties of the model clusters with observations. 
The results of this comparison, together with the results of other studies, 
e.g., of the fluctuations in the 3K--background measured by COBE, can then be 
used to find the parameters of the scenario that is most consistent with all 
data. 
For the best--fitting scenario, a cluster catalogue will be constructed later 
on, similar to the standard CDM catalogue of van Kampen \& Katgert (1997), in 
which each cluster will be simulated individually at higher resolution. 

\par The cosmological parameters that are varied between the different 
scenarios are the shape of the initial power spectrum, its amplitude which 
gives rise to the present {\it r.m.s.} mass fluctuation on scales of 
$8 h^{-1}$ Mpc, $\sigma_{8}$, the density in units of the critical density, 
$\Omega_{0}$, and the Hubble--parameter, $H_{0}$. 
The values of the parameters are chosen such that very often one can compare 
two scenarios that differ by only one parametervalue, so that the influence of 
that specific parameter on the cluster properties can be investigated. 

\par The comparison with observations is done in a companion paper (de Theije, 
van Kampen \& Slijkhuis 1997, hereafter Paper II). 
In that paper, we will try to mimic as closely as possible the observational 
way of defining a cluster, to make a reliable comparison. 

\par The present paper is organized as follows: in Section \ref{methods} we 
describe the N--body simulations, and the identification of groups in the 
simulations. 
In Section \ref{abundances} we give the number densities of clusters in the 
simulation boxes and compare these to the observations. 
In Section \ref{int_prop} we describe the intrinsic properties of the clusters 
in the simulations, viz. their mass, velocity dispersion, peculiar velocity, 
shape, and alignment with the surroundings. 
In Section \ref{parameters} we compare the properties of clusters for 
different parameter values, which allows us to investigate the influence of a 
given parameter on the cluster properties. 
Finally, in Section \ref{conclusions} the main results are summarized and 
discussed. 

\begin{table}
\centering
\caption{Description of the cosmological scenarios that have been examined. 
The first column gives the acronym, which will be used throughout this paper 
to identify a scenario. 
SCDM indicates the Standard CDM scenario with $\Omega_{0} = 1.0$ and $h=0.5$. 
The various LCDM scenarios all have $\Omega_{0} < 1.0$. 
hCDM denotes a scenario with $h=0.3$, and TCDM is a tilted CDM scenario, 
having a slightly tilted initial spectrum. 
The second column gives the shape of the power spectrum. 
The CDM power spectrum is taken from Davis {\it et al.} (1985), the HDM one 
from Bardeen {\it et al.} (1986). 
Column 3 denotes the present--day {\it r.m.s.} mass fluctuation in spheres of 
$8 h^{-1}$ Mpc. 
For the $\Omega_{0}=1.0$ scenarios, the values are the maximum values we 
investigate (see text). 
Column 4 gives the density parameter. 
Column 5 gives the Hubble--constant in units of 100 km s$^{-1}$ Mpc$^{-1}$. 
Column 6 gives the index of the primordial spectrum, where $n=1.0$ denotes the 
Harrison--Zeldovich spectrum and $n=0.8$ corresponds to a slightly tilted 
initial spectrum. 
Column 7 gives the implied present age of the Universe for each scenario.} 
\begin{tabular}{lcccccc}
\hline
scenario & Power    & $\sigma_{8}$ & $\Omega_{0}$ & $h    $ & $n$ & age \\
         & Spectrum &              &              &       & & ($10^{10}$ yr) \\
\hline
SCDM     & CDM      & 0.61         & 1.0          & 0.5     & 1.0 & 1.3 \\
LCDMa    & CDM      & 0.46         & 0.2          & 0.5     & 1.0 & 1.7 \\
LCDMb    & CDM      & 0.90         & 0.2          & 0.5     & 1.0 & 1.7 \\
LCDMc    & CDM      & 0.90         & 0.2          & 1.0     & 1.0 & 0.8 \\
LCDMd    & CDM      & 0.60         & 0.8          & 0.5     & 1.0 & 1.4 \\
$k^{-2}$ & $k^{-2}$ & 0.64         & 1.0          & 0.5     & --2.0 & 1.3 \\
hCDM     & CDM      & 1.00         & 1.0          & 0.3     & 1.0 & 2.2 \\
TCDM     & CDM      & 0.60         & 1.0          & 0.5     & 0.8 & 1.3 \\
HDM      & HDM      & 1.00         & 1.0          & 1.0     & 1.0 & 0.7 \\
\hline
\hline
\end{tabular}
\label{models}
\end{table}


\section{METHODS}
\label{methods}


\subsection{Simulations}
\label{simulations}

The simulations are performed using a P$^{3}$M code (Bertschinger \& Gelb 
1991) in a box of 256$^{3}$ $h^{-3}$ Mpc$^{3}$ with periodic boundary 
conditions and 128$^{3}$ mesh points. 
128$^{3}$ particles are used, each with a mass of $2.22 \times 10^{12} 
\Omega_{0} h^{-1} M_{\odot}$. 
Each simulation took about 25 hours of CPU--time on the Cray C90 of the SARA 
Computing Center in Amsterdam. 
A softening length of 0.2 $h^{-1}$ Mpc (i.e., one tenth of a grid cell) is 
used. 
This value is the same as Frenk {\it et al.} (1990) adopted, and makes a 
comparison between both studies possible. 
The accuracy of the simulations is judged from the Layzer--Irvine cosmic 
energy equation (see Efstathiou {\it et al.} 1985). 
In all simulations, the integration constant of this equation changes less 
than 0.06\% of the current potential energy at any stage of a run. 

\par The parameters that define the simulations are given in Table 
\ref{models}. 
All simulations are done using the same random number seed to set up the 
initial conditions in order to remove cosmic variance from the comparison of 
the various scenarios. 
The initial conditions are evaluated at $z=9$ by means of the Zeldovich 
approximation. 
For the CDM simulations we use the Davis {\it et al.} (1985) power spectrum. 
For the HDM simulation we use the Bardeen {\it et al.} (1986) power spectrum. 
$\Omega_{0}$ is the present--day value of the density parameter, while $h$ 
describes the present--day value of the Hubble--parameter $H_{0}$ in units of 
100 km s$^{-1}$ Mpc$^{-1}$. 

\par In general, $\sigma_{8}$ increases with time, so it is an indication of 
time. 
For scenarios with $\Omega_{0}=1.0$, $\sigma_{8}$ evolves linearly with the 
expansion factor (e.g., Padmanabhan 1993) because the shape of the power 
spectrum is constant in time. 
In such a scenario the particle distribution at an epoch before the present 
time is identical to that at the present time in a scenario with a 
correspondingly smaller value of $\sigma_{8}$. 
Consequently, the values of $\sigma_{8}$ given for the five $\Omega_{0}=1.0$ 
scenarios, are the maximum values of $\sigma_{8}$ that one can probe. 
The $\sigma_{8}$ values in Table \ref{models} are roughly in agreement with 
the findings of Eke, Cole \& Frenk (1996). 
These authors derived values for $\sigma_{8}$ equal to 0.50 $\pm$ 0.04 for 
$\Omega_{0}=1.0$ scenarios, 0.55 $\pm$ 0.04 for $\Omega_{0}=0.8$, and 1.03 
$\pm$ 0.08 for $\Omega_{0} = 0.2$ scenarios, by comparing their N--body 
simulations with the observed number density of clusters as a function of 
X--ray temperature. 
The LCDMa scenario has a normalization similar to the COBE--normalization, 
$\sigma_{8}=0.46$ (Sugiyama 1995). 

\par For the scenarios with $\Omega_{0}<1.0$, the shape of the power spectrum 
does change with time and therefore the simple rescaling of the time 
coordinate in the $\Omega_{0}=1.0$ scenarios is impossible. 
In these scenarios the particle distribution at an epoch before the present 
time is identical to that at the present time in a scenario with a smaller 
value of $\sigma_{8}$ as well as larger values of $\Omega$ and the spectral 
parameter $\Gamma$. 
The latter was introduced by Efstathiou, Bond \& White (1992) and measures 
the spectral shape. 
For CDM scenarios it is equal to $\Omega h$. 
Scenarios with a larger value of $\Gamma$ contain less power on large scales 
and slightly more power on small scales. 
For example, for the LCDMa and LCDMb scenario, the range in $\Gamma$ is 
0.10--0.27, for the LCDMc scenario 0.20--0.53, and for the LCDMd scenario 
0.40--0.47, for expansion factors between 1.00 and 0.22. 

\par $n$ is the index of the primordial spectrum, with the canonical value 
of 1.0 corresponding to the Harrison--Zeldovich spectrum. 
The TCDM scenario has $n=0.8$, i.e., its spectrum is slightly tilted and has 
somewhat more power on larger scales than standard CDM. 

\par The LCDMd scenario is included to investigate the difference between a 
scenario with a value of $\Omega_{0}$ less than but close to 1.0 and a 
scenario with $\Omega_{0}=1.0$. 
Because of the different dynamics between flat and open scenarios, the fact 
that $\Omega_{0}$ is smaller than 1.0 may be more important for the cluster 
properties than how much $\Omega_{0}$ is exactly below 1.0. 
As most studies for $\Omega_{0}<1.0$ so far usually adopt a 'very low' value 
for $\Omega_{0}$, i.e., $\approx 0.3$, this has never been checked thoroughly.

\par The implied present age of the Universe in each of the scenarios is 
given in column 7 of Table \ref{models}. 

\par The expansion factor $a_{\mbox{exp}}$ for which the particle positions 
and velocities are stored are almost always $a_{\mbox{exp}} = $
(0.22, 0.33, 0.44, 0.55, 0.66, 0.79, 0.89, 1.00), except for the SCDM and 
$k^{-2}$ scenarios. 
For the SCDM scenario we have stored $a_{\mbox{exp}} = $ (0.25, 0.33, 0.41, 
0.49, 0.59, 0.67, 0.75, 0.84, 0.90, 1.00) and for the $k^{-2}$ scenario 
$a_{\mbox{exp}} = $ (0.16, 0.24, 0.31, 0.39, 0.47, 0.56, 0.64, 0.71, 0.86, 
1.00). 
The redshift corresponding to a given expansion factor follows from $z = 
a_{\mbox{exp}}^{-1}-1$. 


\subsection{Defining clusters}
\label{group_finding_3D}


\subsubsection{Group finding algorithm}

\par The algorithm that we used to define groups is the 
``friends--of--friends'' algorithm. 
It is described in, e.g., Davis {\it et al.} (1985). 
This algorithm links all particle pairs that are separated by less than a 
fraction $p$ of the mean interparticle distance. 
Each subset of linked particles is then defined as a group. 
The algorithm finds groups that have an overdensity $p^{-3}$ with respect to 
the mean background density. 
Since the overdensity $\delta \rho / \rho $ within the virial radius of a 
cluster is $\approx 180$ (e.g., Padmanabhan 1993), typical values for $p$ 
should be about 0.15--0.20. 
We adopt $p=0.20$. 
The algorithm has the advantage that it produces a unique catalog of groups 
for any $p$, and that it does not make a priori assumptions about the shape of 
the groups. 

\par Only groups containing at least 100 particles are included in the cluster 
list. 
This lower limit is adopted since fewer particles may result in a large 
increase in shot noise in the measurements. 
In the simulations 100 particles represent a mass of $2.22 \times 10^{14} 
h^{-1} \Omega_{0} M_{\odot}$. 
This mass can be compared to the mass within the virialized part of clusters, 
which is about $5 \times 10^{14} h^{-1} M_{\odot}$.


\subsubsection{Cluster definition}
\label{cluster_definition}

For the study of intrinsic properties of the model clusters, all groups with 
a mass larger than $2.22 \times 10^{14} h^{-1} M_{\odot}$ will be considered 
'clusters' from now on. 
These clusters form our {\it mass--limited} cluster catalogue. 
Each cluster then contains at least 100/$\Omega_{0}$ particles. 
This difference in mass resolution between $\Omega_{0}=1.0$ and 
$\Omega_{0}=0.2$ may influence the analyses. 
In Paper II we will address the problem of mass resolution more thoroughly by 
explicitly undersampling the clusters. 
We conclude there that this undersampling does not have any significant 
influence on the cluster properties. 

\par The observed number density of rich ($R \ge 1$) Abell clusters is $8.6 
\times 10^{-6} h^{3}$ Mpc$^{-3}$ (e.g., Mazure {\it et al.} 1996). 
Thus, one expects about 144 rich clusters in our simulation volume. 
This number of 144 will be used when comparing the observed properties of 
the model clusters to real observations (Paper II). 
As in some of the scenarios the number of clusters with a mass above the 
limit is far less than 144, we have chosen to define for each scenario a 
{\it number--selected} cluster catalogue as well, consisting of the 144 most 
massive clusters. 
However, the mass ranges of those catalogues can be quite different. 
For example, in the LCDMb scenario the mass of the least massive cluster in 
the number--selected catalogues is $1.5 \times 10^{14} h^{-1} M_{\odot}$ 
while for the HDM scenario it is $7.7 \times 10^{14} h^{-1} M_{\odot}$. 
The mass of the most massive cluster ranges from $5.8 \times 10^{14} h^{-1} 
M_{\odot}$ to $2.0 \times 10^{15} h^{-1} M_{\odot}$. 

\par
\begin{figure}
\hspace*{0.5cm}
\psfig{figure=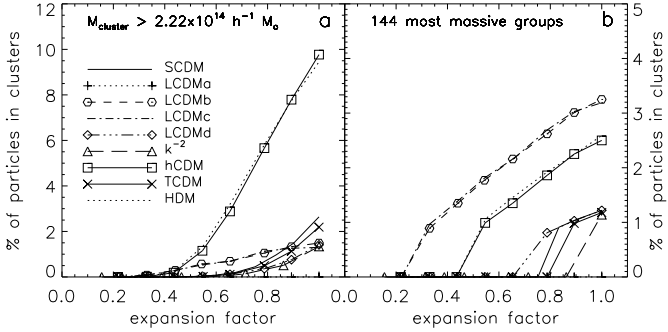,width=7.5cm}
\caption{Percentage of particles in clusters. 
{\bf a}: Clusters are defined as those groups which have a mass of at least 
$2.22 \times 10^{14} h^{-1} M_{\odot}$. 
{\bf b}: Clusters are defined as the 144 most massive groups.}
\label{massper}
\end{figure}

 For both definitions of the cluster sample, Figure \ref{massper} shows 
the percentage of all particles that is in clusters. 
This gives an idea of the importance of such clusters in the simulations. 
For the mass--limited samples, two scenarios clearly stand out, viz. the hCDM 
and HDM scenarios (lefthand panel). 
The reason for this is the large value of $\sigma_{8}$ for these scenarios 
(i.e., $\sigma_{8} = 1.00$, see Table \ref{models}). 
At an expansion factor $a_{\mbox{exp}} \approx 0.65$, where both scenarios 
have $\sigma_{8} \approx 0.65$, the fraction of particles in clusters is 
similar to that in other scenarios which also have $\sigma_{8} \approx 0.65$. 
So, while studying the clusters that are defined in this way, one is treating 
a similar fraction of the total number of particles in each simulation cube. 

\par The righthand panel of Figure \ref{massper} shows the percentage of 
particles in the 144 most massive groups. 
Again, the hCDM and HDM scenarios stand out, but now the two $\Omega_{0}=0.2$ 
scenarios LCDMb and LCDMc have an even larger fraction of the particles in 
clusters, almost three times as much as for the other scenarios. 
If one would rescale these scenarios to lower $\sigma_{8}$ by investigating 
the scenario at earlier times, the density parameter $\Omega$ and the spectral 
parameter $\Gamma$ also change. 
So the fraction of particles involved shows a broader range than for the 
mass--limited samples. 
However, we use the number--selected sample only if the results of the 
mass--limited sample may be influenced by the limited number of clusters in 
the latter sample. 

\begin{table*}
\centering
\caption{Number of clusters with a mass within $r = 1.0 h^{-1}$ Mpc of at 
least $2.22 \times 10^{14} h^{-1} M_{\odot}$, corresponding to 
$100/\Omega_{0}$ particles. 
The numbers in parentheses give the mean number of particles per cluster, 
averaged over all clusters.} 
\begin{tabular}{lrrrrrrrrrr}
\hline
scenario & \multicolumn{10}{c}{expansion factor} \\
\cline{2-11}
      & 0.22 & 0.33 & 0.44 & 0.55 & 0.66 & 0.79 & 0.89 & 1.00 & & \\
\hline
LCDMa & 0 & 0 & 0 & 0 & 0 & 0 & 0 & 0 & & \\
LCDMb & 1 & 5 (319) & 17 (360) & 30 (396) & 29 (501) & 39 (567) & 44 (637) & 46 (679) & & \\
LCDMc & 1 & 5 (316) & 17 (341) & 28 (396) & 32 (461) & 42 (563) & 41 (644) & 41 (662) & & \\
LCDMd & 0 & 0 & 0 & 6 (127) & 18 (138) & 51 (146) & 100 (160) & 181 (164) & & \\
hCDM  & 0 & 2 (108) & 30 (130) & 181 (134) & 424 (143) & 749 (159) & 962 (170) & 1166 (176) & & \\
TCDM  & 0 & 0 & 0 & 3 (108) & 18 (131) & 82 (126) & 180 (132) & 338 (136) & & \\
HDM   & 0 & 2 (121) & 38 (129) & 205 (139) & 455 (151) & 747 (163) & 928 (175) & 1062 (187) & & \\
\hline
\hline
      &   &         &          &           &           &           &         \\
\hline
scenario & \multicolumn{10}{c}{expansion factor} \\
\cline{2-11}
      & 0.25 & 0.33 & 0.41 & 0.49 & 0.59 & 0.67 & 0.75 & 0.84 & 0.90 & 1.00 \\
\hline
SCDM  & 0 & 0 & 0 & 1 (106) & 12 (128) & 34 (122) & 77 (124) & 159 (127) & 232 (134) & 406 (136) \\
\hline
\hline
      &   &         &          &           &           &           &         \\
\hline
scenario & \multicolumn{10}{c}{expansion factor} \\
\cline{2-11}
      & 0.16 & 0.24 & 0.31 & 0.39 & 0.47 & 0.56 & 0.64 & 0.71 & 0.86 & 1.00 \\
\hline
$k^{-2}$ & 0 & 0 & 0 & 0 & 0 & 1 (100) & 8 (131) & 23 (125) & 72 (147) & 189 (149) \\
\hline
\hline
\end{tabular}
\label{tababundances}
\end{table*}


\subsubsection{Cluster centres}
\label{cluster_center}

To determine the cluster centre, two methods are used. 
The first one is the same as in de Theije, Katgert \& van Kampen (1995). 
First, one calculates the centre of mass of all particles in the cluster. 
Then one defines an aperture of radius $1.0 h^{-1}$ Mpc around this centre 
and calculates the new centre of all particles within this aperture. 
This procedure is repeated until the mass centre does not shift by more than 
$0.1 h^{-1}$ Mpc. 

\par In the second approach to define the cluster centre we find the particle 
which has the largest smoothed density of particles around it. 
This density is calculated by smoothing the particle distribution around each 
particle with a Gaussian distribution of dispersion $R_{s}$, which is equal to 
half the average nearest neighbour distance $[3 / 4 \pi \langle n 
\rangle]^{1/3}$, where $\langle n \rangle$ is the mean number density of 
particles (see van Kampen 1995). 
This definition resembles the X--ray centre of a cluster, because the X--ray 
emission is proportional to the electron density squared of the intracluster 
gas. 

\par We have checked that the cluster mass is the same for both definitions of 
cluster centre, and expect this to be true for the other cluster properties as 
well. 
As the first method is much less time consuming, we use that method in what 
follows. 


\subsection{Galaxy identification}
\label{galaxy_identification}

Different schemes have been used in the literature to identify galaxies in 
N--body simulations (e.g., White {\it et al.} 1987, Nolthenius, Klypin \& 
Primack 1994). 
In most of the schemes one identifies galaxies as the highest peaks above a 
threshold given by some bias factor. 
Van Kampen (1995) described the formation and evolution of galaxies by 
replacing a group of particles that is roughly in virial equilibrium by a 
single soft particle with mass, position, velocity and softening corresponding 
to that group. 
His constrained random field single cluster simulation cubes were much smaller 
than the present ones and each particle had a mass of $3.5 \times 10^{10} 
h^{-1} M_{\odot}$. 
As the particles in our simulations have a mass of $ 2.22 \times 10^{12} 
\Omega_{0} h^{-1} M_{\odot}$, already the mass of a galaxy, it is impossible 
to apply this algorithm here. 

\par Instead, we assume the dark matter to be equally distributed as the 
luminous matter, i.e., galaxies and gas. 
Buote \& Canizares (1996) concluded that, for their sample of 5 clusters, 
the shapes of the dark matter distribution, the mass distribution and galaxy 
isopleths are all consistent with each other. 
In addition, for the catalogue of model clusters of van Kampen (1995), the 
distributions of ellipticities for the galaxy particles on the one hand, 
and for all dark matter particles on the other hand are statistically the 
same. 

\par From a sample of 41 clusters with measured velocity dispersions and 
X--ray temperatures, Lubin \& Bahcall (1993) concluded that the velocity bias 
in clusters, $b_{v} \equiv \sigma_{\mbox{gal}}/\sigma_{\mbox{DM}}=0.97 \pm 
0.04$. 
So the velocity dispersion of the galaxies is almost identical to that of the 
dark matter. 
Also, van Kampen (1995) found no evidence for velocity bias in his catalogue 
of cluster models which contain a recipe for galaxy formation. 
In addition, several authors concluded that the spatial distribution of the 
galaxies, gas and the total mass are all very similar, with possibly a 
somewhat larger central concentration for the dark matter (e.g., Henry, Briel 
\& Nulsen 1994, Smail {\it et al.} 1995, Tyson \& Fischer 1995, Squires {\it 
et al.} 1996). 

\begin{figure}
\hspace*{0.7cm}
\psfig{figure=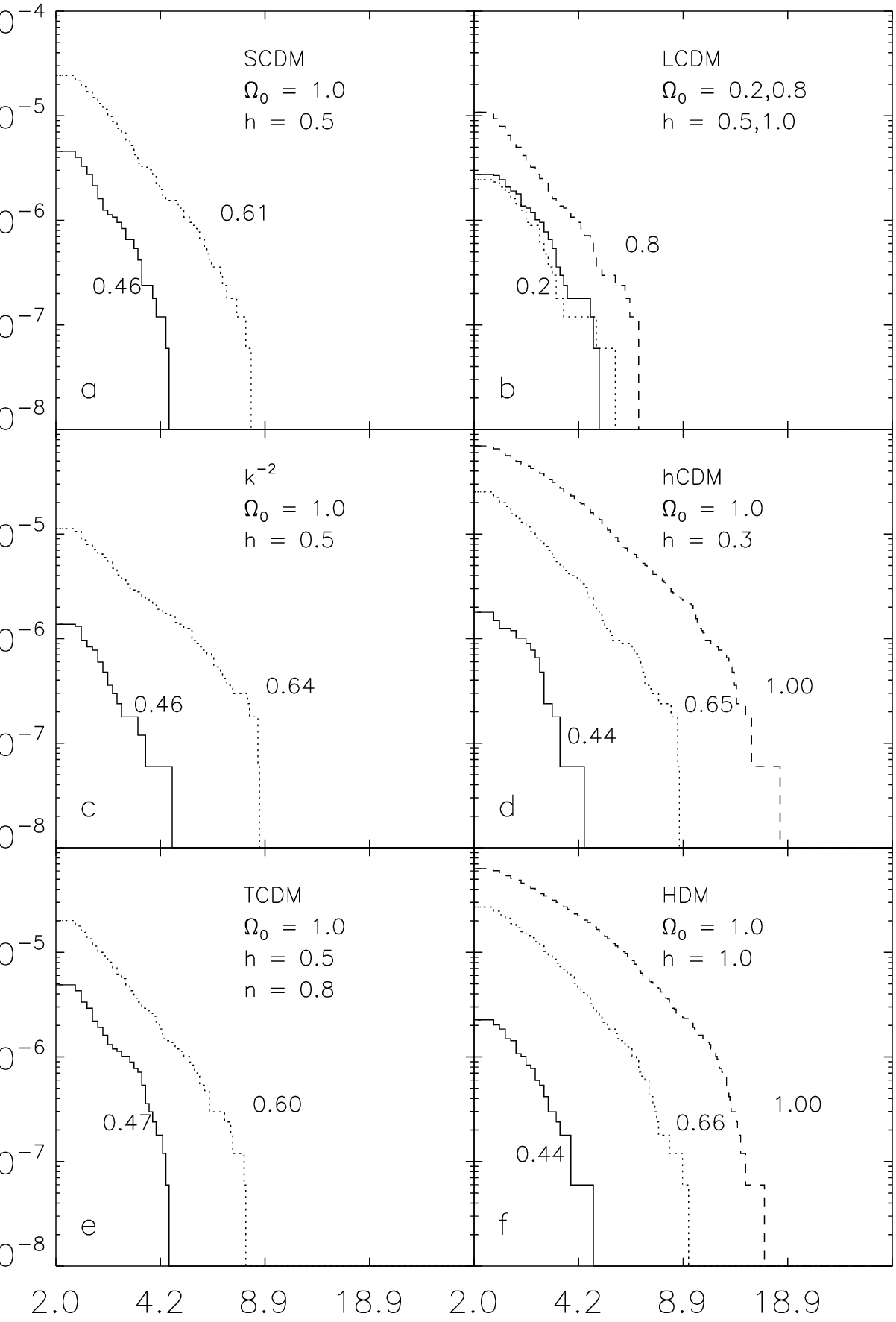,width=7.5cm}
\vspace{0.6cm}
\caption{Cumulative distribution of cluster particle masses for all clusters 
having a particle mass of at least $2.22 \times 10^{14} h^{-1} M_{\odot}$. 
{\bf a}: SCDM with $\sigma_{8}=0.46$ (solid line) and $\sigma_{8}=0.61$ 
(dotted line). 
{\bf b}: LCDMb (solid line), LCDMc (dotted line) and LCDMd (dashed line). 
{\bf c}: $k^{-2}$ with $\sigma_{8}=0.46$ (solid line) and 
$\sigma_{8}=0.64$ (dotted line). 
{\bf d}: hCDM with $\sigma_{8}=0.44$ (solid line), $\sigma_{8}=0.65$ 
(dotted line) and $\sigma_{8}=1.00$ (dashed line). 
{\bf e}: TCDM with $\sigma_{8}=0.47$ (solid line) and $\sigma_{8}=0.60$ 
(dotted line). 
{\bf f}: HDM with $\sigma_{8}=0.44$ (solid line), $\sigma_{8}=0.66$ 
(dotted line) and $\sigma_{8}=1.00$ (dashed line).}
\label{nummass}
\end{figure}


\section{CLUSTER NUMBER DENSITIES}
\label{abundances}

An important property of the cluster distribution as a whole is the number 
density of rich Abell clusters in the simulations. 
In the present paper we define the cluster sample to be all groups in the 
simulation that have a mass of at least $2.22 \times 10^{14} h^{-1} M_{\odot}$ 
(see Section \ref{cluster_definition}), i.e., the mass--limited catalogue. 

\par In Table \ref{tababundances} we give the number of clusters in each 
simulation, i.e., the number of groups which have a mass of at least $2.22 
\times 10^{14} h^{-1} M_{\odot}$ within a clustercentric radius of $1.0 
h^{-1}$ Mpc at various times. 
For the SCDM scenario, these values are similar to the predicted number of 
clusters above this minimum mass as derived from the Press--Schechter 
formalism (Eke {\it et al.} 1996). 
The numbers in parentheses in Table \ref{tababundances} give the mean number 
of particles per cluster within $1.0 h^{-1}$ Mpc, averaged over all clusters. 
This mean number of particles increases towards later times for all 
cosmological scenarios, illustrating the growth of the clusters. 

\par In the LCDMa scenario, no clusters with $M \ge 2.22 \times 10^{14} h^{-1} 
M_{\odot}$ are formed for the assumed value $\sigma_{8}=0.46$ in the 
simulation. 
Therefore, we will not consider this scenario any further. 

\par If one evaluates the {\it total} mass of the clusters, using the entire 
group as defined by the friends--of--friends algorithm, the results are 
qualitatively the same as when one uses the mass within $1.0 h^{-1}$ Mpc. 


\section{INTRINSIC PROPERTIES OF CLUSTERS}
\label{int_prop}

The intrinsic properties of the model clusters are evaluated using both the 
mass--limited and the number--selected cluster catalogues. 
In the following analyses, all cluster particles that are within a 
clustercentric radius of $r = 1.0 h^{-1}$ Mpc are considered, unless stated 
otherwise. 


\subsection{Masses}
\label{mass}

The first, most basic, property of the clusters that we will investigate is 
their mass $M$. 
Figure \ref{nummass} shows the cumulative distribution $\rho(>M)$ of particle 
mass for all clusters in the mass--limited sample. 
For the $\Omega_{0}=1.0$ scenarios, different values of $\sigma_{8}$ are 
plotted. 
In these cases, $\sigma_{8}$ is highest for the upper curves and decreases 
downward. 
In the upper righthand panel, the low--$\Omega_{0}$ scenarios are shown. 
For the $\Omega_{0}=1.0$ scenarios, the cluster mass increases with 
$\sigma_{8}$, or, equivalently, with time. 
If one were to extrapolate this to $\sigma_{8}=0.9$, the value adopted for the 
LCDMb and LCDMc scenarios, the clusters in the $\Omega_{0}=1.0$ scenarios have 
a much larger mass than those in these two low--$\Omega_{0}$ scenarios for the 
same value of $\sigma_{8}$. 
Even in the LCDMd scenario with $\Omega_{0}=0.8$, the cluster mass is 
significantly smaller than in the SCDM scenario with $\sigma_{8}=0.61$. 

\par A comparison of the distributions for the LCDMb and LCDMc scenarios is a 
check of the calculations, as these scenarios are exactly the same apart from 
the value of the Hubble parameter $h$. 
As the masses here are expressed in units of $10^{14} h^{-1} M_{\odot}$, there 
should be no difference at all between these two scenarios for their cluster 
masses. 
Indeed, we find very good agreement. 

\par Among the different scenarios, the particle mass of clusters may be 
quite different. 
The low--$\Omega_{0}$ scenarios have a most massive cluster of about $6 
\times 10^{14} h^{-1} M_{\odot}$. 
The SCDM, $k^{-2}$ and TCDM scenarios contain masses up to $\approx 8 \times 
10^{14} h^{-1} M_{\odot}$. 
The hCDM and HDM scenarios contain clusters of even larger mass, $18 \times 
10^{14} h^{-1} M_{\odot}$. 
This is due to the high value of $\sigma_{8}$, viz. $\sigma_{8}=1.00$, in 
these latter two scenarios. 

\begin{table}
\centering
\caption{Evolution parameters of the mass of all clusters having a mass of at 
least $2.22 \times 10^{14} h^{-1} M_{\odot}$. 
The values in this table denote the median values of the cluster particle mass 
$M$ (in units of $10^{14} h^{-1} M_{\odot}$). 
The information is listed only for expansion factors at which there are at 
least 10 clusters.}
\begin{tabular}{lcccccc}
\multicolumn{7}{c}{median value of cluster particle mass (in $10^{14} h^{-1} 
M_{\odot}$)} \\
\hline
scenario & \multicolumn{6}{c}{expansion factor} \\
\cline{2-7}
      & 0.44 & 0.55 & 0.66 & 0.79 & 0.89 & 1.00 \\
\hline
LCDMb &     &     & 2.9 & 2.9 & 3.0 & 2.9 \\
LCDMc &     &     &     & 2.8 & 3.1 & 2.8 \\
LCDMd &     &     & 2.6 & 2.5 & 2.6 & 2.7 \\
hCDM  & 2.9 & 2.7 & 2.8 & 3.1 & 3.2 & 3.3 \\
TCDM  &     &     & 2.7 & 2.6 & 2.7 & 2.7 \\
HDM   & 2.7 & 2.8 & 3.0 & 3.2 & 3.3 & 3.5 \\
\hline
\hline
      &     &     &     &     &     &     \\
\hline
scenario & \multicolumn{6}{c}{expansion factor} \\
\cline{2-7}
      & 0.59 & 0.67 & 0.75 & 0.84 & 0.90 & 1.00 \\
\hline
SCDM  & 2.8 & 2.7 & 2.6 & 2.6 & 2.8 & 2.8 \\
\hline
\hline
      &     &     &     &     &     &     \\
\hline
scenario & \multicolumn{6}{c}{expansion factor} \\
\cline{2-7}
      & 0.47 & 0.56 & 0.64 & 0.71 & 0.86 & 1.00 \\
\hline
$k^{-2}$ &  &     &     & 2.7 & 3.0 & 2.8 \\
\hline
\hline
\end{tabular}
\label{tabevol1}
\end{table}

\begin{figure}
\hspace*{1cm}
\psfig{figure=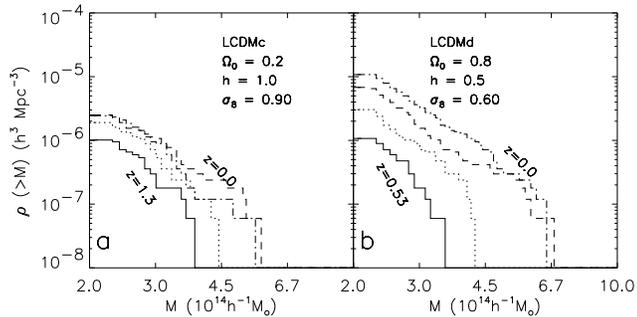,width=7cm}
\caption{Time evolution of the cumulative distribution of cluster particle 
mass.
Only clusters with a mass of at least $2.22 \times 10^{14} h^{-1} M_{\odot}$
are considered.
{\bf a}: LCDMc scenario.
The curves correspond to redshifts $z=1.3$ (solid line), 0.53 (dotted line),
0.27 (dashed line), and 0.0 (dot--dashed line).
{\bf b}: LCDMd scenario.
The curves correspond to redshifts $z=0.53$ (solid line), 0.27 (dotted line),
0.12 (dashed line), and 0.0 (dot--dashed line).}
\label{lowOmegaevol2}
\end{figure}

\par For all $\Omega_{0}=1.0$ scenarios the slope of $\rho(>M)$ is clearly 
flatter at later times, showing that the relative number of high--mass 
clusters increases at later times. 
The change in the median value of the cluster particle mass, however, is not 
very large. 
This is clear from Table \ref{tabevol1}, which shows the median value of the 
cluster particle mass for different values of $a_{\mbox{exp}}$. 
Only for the hCDM and HDM scenarios is there an increase in the median value 
towards later times. 
For all other scenarios the increase is not significant or not monotonic. 

\par The median value of the cluster particle mass in the low--$\Omega_{0}$ 
scenarios changes less rapidly with time. 
To illustrate the change of cluster mass in these open scenarios, we show in 
Figure \ref{lowOmegaevol2} the change of the particle mass distribution of the 
low--$\Omega_{0}$ scenarios LCDMc and LCDMd with time. 
For the LCDMc scenario, the cluster population does not evolve strongly for 
redshifts $z \le 0.50$. 
I.e., the total number of clusters is more or less constant and only a few 
high--mass clusters increase their mass even further. 
This result is quantitatively consistent with that of, e.g., Eke {\it et al.} 
(1996). 
In the LCDMd scenario, the situation is quite different. 
Here the number of clusters increases up to the present time and the maximum 
mass increases somewhat faster than in the LCDMc scenario. 
So there is a clear difference in the time--evolution of cluster mass with 
$\Omega_{0}$, even between scenarios with $\Omega_{0} < 1.0$. 
Alternatively, the different curves in Figure \ref{lowOmegaevol2} can also 
be viewed as corresponding to scenarios with a different value of $\Gamma$ 
(see Section \ref{simulations}). 
In that case, $\Gamma$ is lowest for the upper curves. 
Scenarios with a larger value of $\Gamma$ then seem to have less clusters 
above a certain mass, but this result is solely due to the different values of 
$\sigma_{8}$ that correspond to the different curves. 

\par If one uses the average mass estimator (Heisler, Tremaine \& Bahcall 
1985)
\begin{equation}
M_{\mbox{ave}} = \frac{5.6}{GN(N-1)}\sum_{i<j} (v_{zi}-v_{zj})^{2}R_{ij},
\end{equation}
where $G$ is the gravitational constant, $N$ the total number of particles, 
$v_{zi}$ is the velocity of galaxy $i$ w.r.t. the cluster center, and $R_{ij}$ 
is the distance between galaxies $i$ and $j$, then the cumulative 
distributions of cluster masses look very similar to the distributions of 
cluster particle mass. 
The average mass estimator, however, overpredicts the cluster particle mass by 
about 40\% in all scenarios. 
The projected mass estimator overpredicts the cluster particle mass even more 
than the average mass estimator. 
These results are quantitatively similar to those of van Kampen (1995), who 
used N--body simulations of single clusters in an $\Omega_{0}=1.0$ CDM 
Universe to test the mass estimators. 
He found that the average mass estimator overpredicts the cluster mass by 
about 40\%, if evaluated within about $1.0 h^{-1}$ Mpc, and that the projected 
mass estimator overpredicts the cluster mass by about 60\%. 

\par To investigate how the mass increases with clustercentric radius, we 
determine the mass of the clusters within an Abell radius, $1.5 h^{-1}$ Mpc. 
Using the particle mass for this, the mass within the Abell radius is about 
25--35\% larger than that within $1.0 h^{-1}$ Mpc. 
This is more or less what is expected for a spherical cluster whose density 
profile follows a modified Hubble law $ \rho(r) = \rho(0) / 
[1+(r/r_{c})^{2}]^{3/2} $ with a core radius of $r_{c} = 0.25 
h^{-1}$ Mpc (e.g., Sarazin 1986). 

\par If one uses the number--selected cluster samples, the cumulative 
distributions of cluster mass would be very similar to the curves in Figure 
\ref{nummass} of the mass--limited samples. 
The distributions of cluster average mass change in the same way as those over 
cluster particle mass, but the transition between $\rho(>M) \le 8.6 \times 
10^{-6} h^{3}$ Mpc$^{-3}$ and $\rho(>M) \ge 8.6 \times 10^{-6} h^{3}$ 
Mpc$^{-3}$ is now rather smooth because of the scatter in the relation 
between a cluster's particle mass and average mass. 

\begin{figure}
\hspace*{0.7cm}
\psfig{figure=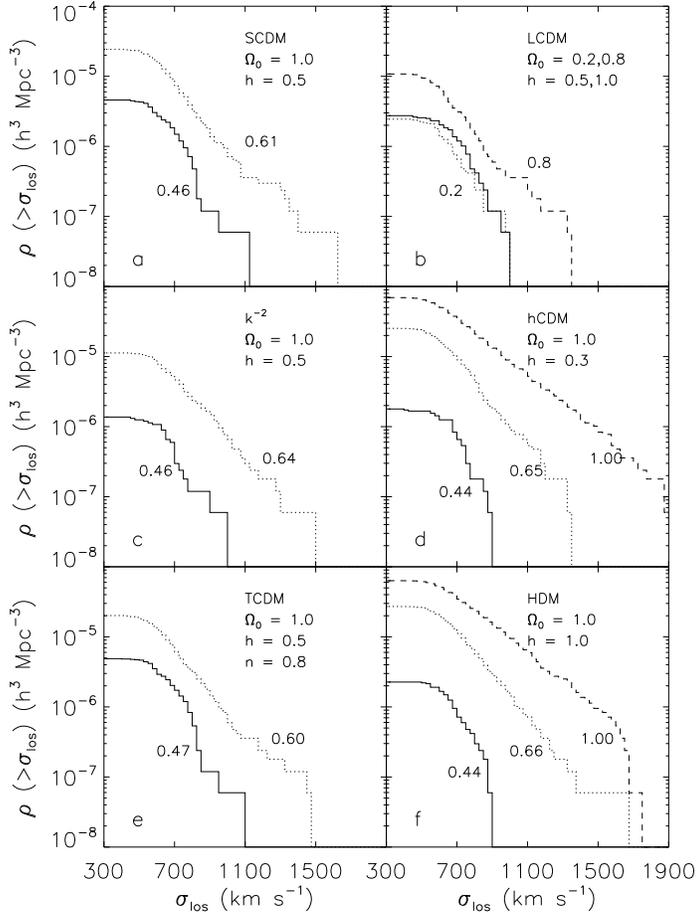,width=7.5cm}
\vspace{0.6cm}
\caption{Cumulative distribution of line--of--sight velocity dispersion for 
clusters in the different cosmological scenarios. 
Only clusters with a mass of at least $2.22 \times 10^{14} h^{-1} M_{\odot}$ 
are considered. 
The different panels and curves have the same meaning as in Figure 2.}
\label{veldispfig}
\end{figure}

\begin{table}
\centering
\caption{Evolution parameters of the line--of--sight velocity dispersion of 
all clusters having a mass of at least $2.22 \times 10^{14} h^{-1} M_{\odot}$. 
The values in this table denote the median values of the line--of--sight 
velocity dispersion $\sigma_{\mbox{los}}$ (in km s$^{-1}$). 
The information is listed only for expansion factors at which there at least 
10 clusters.}
\begin{tabular}{lcccccc}
\multicolumn{7}{c}{median value of cluster line--of--sight velocity dispersion} \\
\hline
scenario & \multicolumn{6}{c}{expansion factor} \\
\cline{2-7}
      & 0.44 & 0.55 & 0.66 & 0.79 & 0.89 & 1.00 \\
\hline
LCDMb &     &     & 652 & 665 & 672 & 683 \\
LCDMc &     &     &     & 654 & 657 & 635 \\
LCDMd &     &     & 625 & 661 & 646 & 619 \\
hCDM  & 670 & 638 & 642 & 661 & 679 & 690 \\
TCDM  &     &     & 634 & 654 & 628 & 638 \\
HDM   & 663 & 649 & 669 & 686 & 702 & 719 \\
\hline
\hline
      &     &     &     &     &     &     \\
scenario & \multicolumn{6}{c}{expansion factor} \\
\cline{2-7}
      & 0.59 & 0.67 & 0.75 & 0.84 & 0.90 & 1.00 \\
\hline
SCDM  & 568 & 650 & 645 & 632 & 638 & 637 \\
\hline
\hline
      &     &     &     &     &     &     \\
\hline
scenario & \multicolumn{6}{c}{expansion factor} \\
\cline{2-7}
      & 0.47 & 0.56 & 0.64 & 0.71 & 0.86 & 1.00 \\
\hline
$k^{-2}$ &     &     &     & 637 & 679 & 665 \\
\hline
\hline
\end{tabular}
\label{tabevol2}
\end{table}


\subsection{Line--of--sight velocity dispersions}
\label{veldisp}

The next property of clusters that we consider is their line--of--sight 
velocity dispersion $\sigma_{\mbox{los}}$, which describes the dynamical state 
of the cluster, as it is influenced by merging events, substructure and the 
shape of the galaxy orbits. 

\par Crone \& Geller (1995) studied the evolution of the cluster velocity 
dispersion using large--scale N--body simulations. 
They concluded that $\sigma_{\mbox{los}}$ can be significantly influenced by 
merger activity and therefore does not simply reflect the cluster mass (they 
found a scatter of about 5--10\% in $\sigma_{\mbox{los}}$ for clusters of the 
same mass). 
Furthermore, they detected some change in $\sigma_{\mbox{los}}$ with time. 
The slope of the cumulative distribution of the number density of clusters 
with a velocity dispersion larger than $\sigma_{\mbox{los}}$, 
$\rho(>\sigma_{\mbox{los}})$, flattens with time, so that at later times more 
clusters are found with large $\sigma_{\mbox{los}}$. 
The change with time is most evident for scenarios with a high value of 
$\Omega_{0}$. 

\par The velocity dispersion for the clusters in our N--body simulations is 
calculated using the robust biweight estimator of Beers, Flynn \& Gebhardt 
(1990), yet the velocity dispersions are essentially the same with the 
ordinary definition. 
The cumulative distributions of line--of--sight velocity dispersion found in 
our simulations are plotted in Figure \ref{veldispfig}. 
As before, these distributions include all clusters which have a mass of at 
least $2.22 \times 10^{14} h^{-1} M_{\odot}$. 
The different panels and curves have the same meaning as in Figure 
\ref{nummass}, i.e., for the $\Omega_{0}=1.0$ scenarios the higher curves 
correspond to larger values of $\sigma_{8}$. 

\begin{figure}
\hspace*{1cm}
\psfig{figure=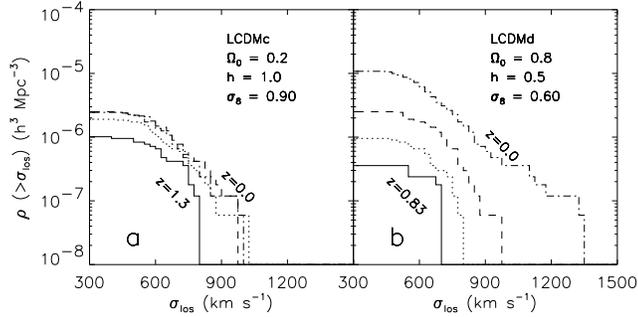,width=7cm}
\caption{Time evolution of the cumulative distribution of cluster velocity 
dispersions. 
Only clusters with a mass of at least $2.22 \times 10^{14} h^{-1} M_{\odot}$ 
are considered. 
{\bf a}: LCDMc scenario. 
The curves correspond to redshifts $z=1.3$ (solid line), 0.53 (dotted line), 
0.27 (dashed line), and 0.0 (dot--dashed line). 
{\bf b}: LCDMd scenario. 
The curves correspond to redshifts $z=0.83$ (solid line), 0.53 (dotted line),
0.27 (dashed line), and 0.0 (dot--dashed line).}
\label{lowOmegaevol}
\end{figure}

\par For the $\Omega_{0}=1.0$ scenarios, our results are consistent with those 
of Jing \& Fang (1994) and Crone \& Geller (1995). 
That is, the curves for lower $\sigma_{8}$, or, alternatively, higher $z$ 
(see Section \ref{simulations}), are steeper than the $z=0$ curves (the upper 
ones). 
Clearly, the evolution of velocity dispersions is not self--similar: the curve 
shifts towards higher $\sigma_{\mbox{los}}$, but at the same time it flattens. 
At the present epoch, clusters with a high $\sigma_{\mbox{los}}$ are 
relatively more abundant than clusters with a low value of 
$\sigma_{\mbox{los}}$. 
These changes in time are analogous to that for the mass of the clusters 
(see the previous section). 
Table \ref{tabevol2} lists the median values of $\sigma_{\mbox{los}}$ for the 
cluster sample at all time steps and for all scenarios for which there is a 
significant number of clusters present that satisfy the minimum mass 
criterion. 
The scenarios with a low $\sigma_{8}$ have relatively smaller values of 
$\sigma_{\mbox{los}}$. 
This is because less matter has collapsed on cluster scales. 
On the other hand, clusters that are not in the sample at early times, because 
their mass is too low, can enter the sample at later times and populate the 
low--$\sigma_{\mbox{los}}$ part of the distribution. 
The combination of both facts results in an apparent upward shift of the whole 
distribution. 
Only the hCDM and HDM scenarios show a significant increase in the median 
value of $\sigma_{\mbox{los}}$ with time. 
For both scenarios we already concluded that the median values of the cluster 
particle mass increase with time (see Section \ref{mass}). 

\par As an example of the evolution of the velocity dispersions in 
low--$\Omega_{0}$ scenarios, we show $\rho(>\sigma_{\mbox{los}})$ for the 
LCDMc and LCDMd scenarios in Figure \ref{lowOmegaevol}. 
For the LCDMc scenario, $\rho(>\sigma_{\mbox{los}})$ hardly changes for $z \le 
0.50$. 
The number of high--$\sigma_{\mbox{los}}$ clusters remains constant, as does 
the maximum value of $\sigma_{\mbox{los}}$. 
This is consistent with results of Jing \& Fang (1994) who found no evolution 
in $\rho(>\sigma_{\mbox{los}})$ for $z \le 0.5$ for their sample of clusters 
with mass larger than $1.7 \times 10^{13} M_{\odot}$. 
Note that these authors used $\Omega_{0}=0.3$. 
In addition, $\sigma_{\mbox{los}}$ does not exceed 1000 km s$^{-1}$, in 
contrast to most of the $\Omega_{0}=1.0$ scenarios. 
For the LCDMd scenario the situation is quite different. 
The number of clusters rises until the present time. 
The median value of $\sigma_{\mbox{los}}$, however, is fairly constant. 
For the different values of $\Omega_{0}$, $\rho(>\sigma_{\mbox{los}})$ changes 
with time in a similar manner as $\rho(>M)$ (see Section \ref{mass}). 

\begin{figure}
\hspace*{1cm}
\psfig{figure=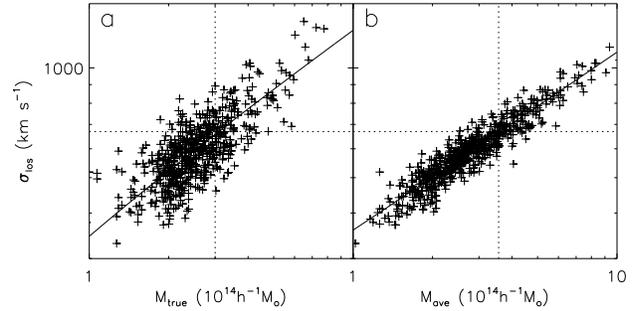,width=7cm}
\caption{Scatter plot of the cluster line--of--sight velocity dispersion 
$\sigma_{\mbox{los}}$ versus its mass for clusters in the SCDM scenario. 
The solid line shows the relation $\sigma \propto \sqrt{M_{\mbox{ave}}}$. 
The horizontal and vertical dotted lines indicate the values of 
$\sigma_{\mbox{los}}$ and $M$, respectively, of the 144--th most massive 
group, in terms of the particle mass. 
{\bf a}: $\sigma_{\mbox{los}}$ versus cluster particle mass. 
{\bf b}: $\sigma_{\mbox{los}}$ versus cluster average mass estimator 
$M_{\mbox{ave}}$.}
\label{sigma_m}
\end{figure}

\par For the hCDM and HDM scenarios the fractional change in mass is, in 
general, about twice as large as for the velocity dispersion (Tables 
\ref{tabevol1} and \ref{tabevol2}), consistent with the virial theorem 
estimates. 
This evolutionary difference between mass and velocity dispersion is also 
consistent with the findings of Crone \& Geller (1995), who found that 
the velocity dispersion evolves less rapidly than the mass of a cluster. 
They attributed this to two processes: 
first, for a specific mass range, the velocity dispersion decreases with time 
due to relaxation. 
Secondly, mergers will make the distribution of $\sigma_{\mbox{los}}$ more 
irregular than that over mass. 
That is, the scatter in the $\sigma_{\mbox{los}}-M$--relation due to mergers 
and accretions introduces random fluctuations in $\rho(>\sigma_{\mbox{los}})$. 

\par To check how large the scatter between $\sigma_{\mbox{los}}$ and $M$ is, 
we show in the lefthand panel of Figure \ref{sigma_m} the line--of--sight 
velocity dispersion versus particle mass for all clusters in the SCDM scenario 
for $\sigma_{8}=0.61$. 
The solid line shows, for comparison, the best--fitting relation 
$\sigma_{\mbox{los}} \propto \sqrt{M}$, as expected for systems in virial 
equilibrium. 
The {\it r.m.s.} scatter about this relation is about 14\%, somewhat larger 
than the 5--10\% that Crone \& Geller (1995) found. 
These results are similar for all scenarios. 
For comparison, the righthand panel of Figure \ref{sigma_m} shows the same 
scatter plot but using the average mass estimator. 
The scatter about the linear relation $\sigma_{\mbox{los}} \propto 
\sqrt{M_{\mbox{ave}}}$ is smaller than if using the cluster particle mass, 
namely 7\%. 
So although the average mass estimator overestimates the cluster particle mass 
by about 40\% (see Section \ref{mass}), it correlates better with the cluster 
line--of--sight velocity dispersion than does the cluster particle mass. 
This is because both the velocity dispersion and the average mass estimator 
explicitly use the particle velocities whereas the particle mass does not 
contain this information. 

\par The horizontal and vertical dotted lines in Figure \ref{sigma_m} indicate 
the values of $\sigma_{\mbox{los}}$ and $M$, respectively, of the 144--th most 
massive cluster, in terms of the particle mass. 
These values will be called $\sigma_{144}$ and $M_{144}$ from here on. 
This comparison gives an idea of how complete, in terms of mass, a sample of 
clusters will be that is selected on the basis of their velocity dispersion. 
This is the case, e.g., for the ENACS--survey (Katgert {\it et al.} 1996), 
that is claimed to be complete for $\sigma_{\mbox{los}} \ge 800$ km s$^{-1}$. 
Remember that we expect 144 clusters in a simulation box of (256 $h^{-1}$ 
Mpc)$^{3}$ on the basis of the cluster number density of the ENACS--survey. 
When one uses the cluster particle mass, the fraction of clusters with 
$\sigma_{\mbox{los}} \ge \sigma_{144}$ that also have $M \ge M_{144}$ ranges 
from 64\% (SCDM) to 78\% (for the $k^{-2}$ scenario). 
These numbers are comparable to the 66\% of clusters of richness class $R \ge 
1$ in the Abell catalogue that also has the required intrinsic richness (van 
Haarlem, Frenk \& White 1997). 

\par To check if the velocity dispersion changes with clustercentric radius, 
we calculate the value of $\sigma_{\mbox{los}}$ within a clustercentric 
radius $r = 1.5 h^{-1}$ Mpc. 
The velocity dispersions within the canonical Abell radius of $1.5 h^{-1}$ Mpc 
are, on average, 5\% smaller than those within $1.0 h^{-1}$ Mpc. 
The exact numbers range from 2\% to 7\% and depend on the scenario and on the 
value of $\sigma_{8}$. 

\par For the number--selected cluster samples, the curves of 
$\rho(>\sigma_{\mbox{los}})$ would be very similar to those of the 
mass--limited samples. 
For those scenarios that have at least 144 clusters in the mass--limited 
samples, the curve for the number--selected sample is equal to that of the 
mass--limited sample for $\rho(>\sigma_{\mbox{los}}) \le 8.6 \times 10^{-6} 
h^{3}$ Mpc$^{-3}$, and flattens off smoothly to this constant value for lower 
$\sigma_{\mbox{los}}$. 
The transition is smooth because of the scatter in the 
$\sigma_{\mbox{los}}-M$--relation (see Figure \ref{sigma_m}). 
For scenarios that have less than 144 clusters in the mass--limited sample, 
the curve for the number--selected sample has the same slope for high 
$\sigma_{\mbox{los}}$ and extends to lower $\sigma_{\mbox{los}}$ until it 
flattens off towards $\rho(>\sigma_{\mbox{los}}) = 8.6 \times 10^{-6} h^{3}$ 
Mpc$^{-3}$. 

\begin{figure}
\hspace*{0.7cm}
\psfig{figure=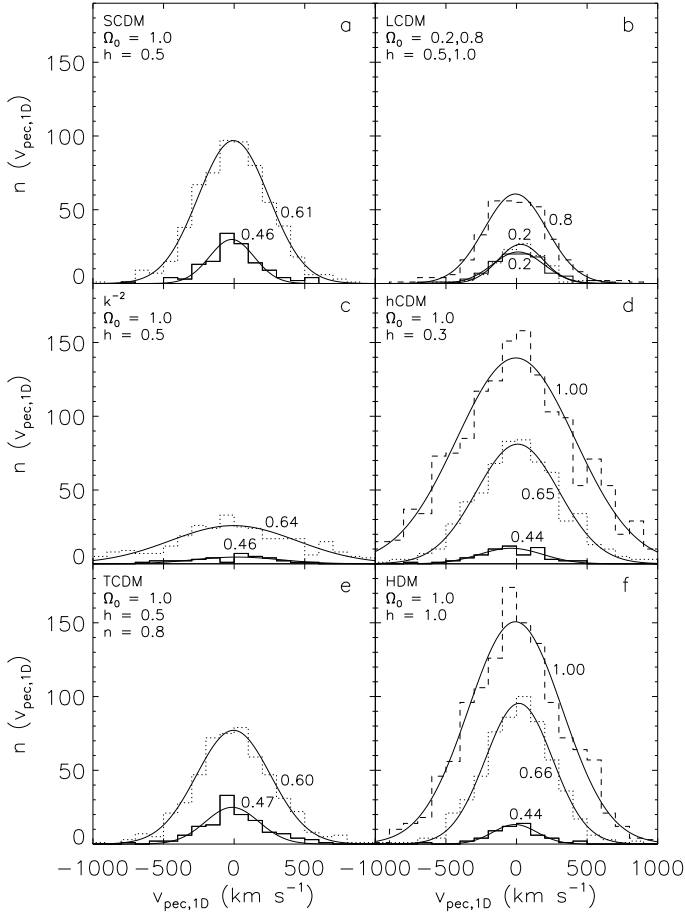,width=7.5cm}
\vspace{0.6cm}
\caption{Distributions of 1--Dl peculiar velocities of clusters in the 
different cosmological scenarios. 
The different panels and curves have the same meaning as in Figure 2. 
The solid lines show the best--fitting Gaussian distributions. 
The parameters of these are listed in Table 4.}
\label{figpecvel}
\end{figure}

\begin{table}
\centering
\caption{Parameter values of the Gaussian fits to the distributions of cluster 
peculiar velocities. 
The first two columns specify the scenario. 
The third and fourth columns describe the best--fitting Gaussian 
distribution, $f(v_{\mbox{pec,1D}}) \propto \exp(-(v_{\mbox{pec,1D}}-
v_{\mbox{pec,1D,0}})^{2}/2 \sigma_{\mbox{pec,1D}}^{2})$. 
The fifth column gives the dispersion of the Maxwellian fits to the 
distribution of 3--D cluster peculiar velocities. 
Column 6 gives the predictions of BBKS using linear theory. 
See text for more details.}
\begin{tabular}{lccccc}
\hline
scenario & $\sigma_{8}$ & $v_{\mbox{pec,1D,0}}$ & $\sigma_{\mbox{pec,1D}}$ & 
$\sigma_{\mbox{pec,3D}}$ & $\sigma_{\mbox{pec,1D,lin}}$ \\
 & & (km s$^{-1}$)  & (km s$^{-1}$) & (km s$^{-1}$) & (km s$^{-1}$) \\
\hline
SCDM     & 0.46      & -16.7          & 159               & 201 & 125 \\
SCDM     & 0.61      &  -4.4          & 251               & 267 & 166 \\
         &           &                &                   &     &     \\
LCDMb    & 0.90      &  12.0          & 182               & 216 & 232 \\
LCDMc    & 0.90      &  29.4          & 160               & 181 & 156 \\
LCDMd    & 0.60      &  -7.3          & 223               & 247 & 162 \\
         &           &                &                   &     &     \\
$k^{-2}$ & 0.46      &  22.9          & 308               & 309 & --- \\
$k^{-2}$ & 0.64      &  -4.6          & 447               & 508 & --- \\
         &           &                &                   &     &     \\
hCDM     & 0.44      & -39.3          & 225               & 242 & 158 \\
hCDM     & 0.65      &   8.9          & 292               & 318 & 234 \\
hCDM     & 1.00      &  -4.6          & 419               & 438 & 360 \\
         &           &                &                   &     &     \\
TCDM     & 0.47      & -14.0          & 190               & 230 & 145 \\
TCDM     & 0.60      &  -4.2          & 262               & 280 & 186 \\
         &           &                &                   &     &     \\
HDM      & 0.44      & -11.1          & 157               & 184 &  97 \\
HDM      & 0.66      & -18.0          & 238               & 254 & 145 \\
HDM      & 1.00      &  -8.0          & 333               & 341 & 221 \\
\hline
\hline
\end{tabular}
\label{tabpecvel}
\end{table}


\subsection{Peculiar velocities}
\label{pecvel}

The next property of galaxy clusters that we consider is the distribution of 
cluster peculiar velocities, i.e., the velocity of a cluster with respect to 
the Hubble--flow. 
This peculiar velocity is calculated using the robust biweight estimator of 
Beers {\it et al.} (1990). 

\par Figure \ref{figpecvel} shows the differential distributions of 1--D 
peculiar velocity $v_{\mbox{pec,1D}}$ of all clusters in the mass--limited 
sample in all scenarios. 
Also shown are the best--fitting Gaussian distributions. 
Such Gaussian distributions are expected for Gaussian random fields (eq. 4.23 
of Bardeen {\it et al.} 1986, hereafter BBKS). 
In general, the Gaussian distributions provide good fits. 
The fitting parameters are given in columns 3 and 4 of Table \ref{tabpecvel}. 
The mean value for the best--fitting Gaussian distributions always differ from 
zero by less than 40 km s$^{-1}$. 
This is expected because there is neither any preferred direction in the 
simulation box nor any systematic flows on the scale of the simulation box. 
The dispersion of the fitted Gaussian distributions increases with 
$\sigma_{8}$, or equivalently cosmic time, for the $\Omega_{0}=1.0$ scenarios. 

\par For a specific value of $\sigma_{8}$, the dispersion of the fitted 
Gaussian distribution is smaller for the $\Omega_{0}<1.0$ scenarios than for 
the $\Omega_{0}=1.0$ scenarios, consistent with earlier findings of, e.g., 
Bahcall, Gramann \& Cen (1994). 
Again, this is because in the open scenarios less matter has collapsed on 
large scales and the total gravitational force acting on a cluster is smaller. 

\par In order to check if the cluster peculiar velocities are distributed 
isotropically, we investigate the distribution of 3--D peculiar velocities, 
$v_{\mbox{pec,3D}}$. 
If the cluster peculiar velocities are distributed isotropically, the 
3--D peculiar velocity vectors should have a random orientation. 
We check this by determining the distributions of spherical angles $\phi$ 
and $\cos(\theta)$ of the velocity vector. 
These distributions are indeed consistent with uniform distributions for all 
scenarios. 
Column 5 of Table \ref{tabpecvel} lists the dispersions 
$\sigma_{\mbox{pec,3D}}$ of the Maxwellian fits $f(v_{\mbox{pec,3D}}) \propto 
v^{2}_{\mbox{pec,3D}} \exp(-(v_{\mbox{pec,3D}}-v_{\mbox{pec,3D,0}})^{2}/2 
\sigma^{2}_{\mbox{pec,3D}})$ to the distributions of 3--D cluster peculiar 
velocity. 
Column 6 gives the expected 1--D dispersions using linear theory (BBKS) and 
peaks on a scale of $4 h^{-1}$ Mpc. 
For the $k^{-2}$ scenario, these values cannot be obtained because the 
integral to calculate $\sigma_{pec,3D}$ diverges (see e.g. Padmanabhan 1993). 
The values of $\sigma_{\mbox{pec,3D}}$ for the model clusters are almost 
always larger than those of the peaks in linear theory. 
This indicates that clusters have evolved into the non--linear regime in most 
of the scenarios. 

\begin{figure}
\hspace*{0.7cm}
\psfig{figure=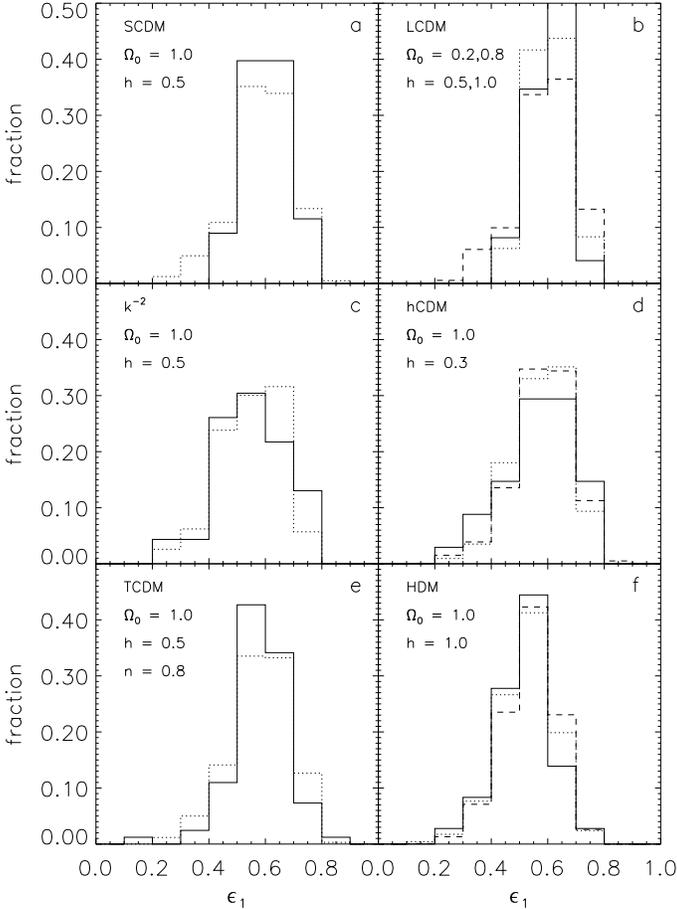,width=7.5cm}
\vspace{0.6cm}
\caption{Distribution of intrinsic cluster ellipticities $\epsilon_{1}$ for 
different cosmological scenarios. 
$\epsilon_{1} = 0.0$ means that the cluster is spherical, $\epsilon_{1} = 1.0$ 
means that the particle distribution of the cluster is a disk. 
Only clusters with a mass of at least $2.22 \times 10^{14} h^{-1} M_{\odot}$ 
are taken into account. 
The different panels and curves have the same meaning as in Figure 2.}
\label{nbodyell}
\end{figure}


\subsection{Shapes}
\label{ell}

De Theije {\it et al.} (1995) used the results of N--body simulations of van 
Kampen (1994) to study how the shapes of clusters may depend on cosmological 
scenario. 
They found that in a low--$\Omega_{0}$ CDM Universe, clusters are, on average, 
more nearly spherical than the {\it same} clusters in an $\Omega_{0}=1.0$ 
Universe. 
This conclusion was also reached by Mohr {\it et al.} (1995) who determined 
cluster morphologies from SPH--simulations. 
These calculations showed that the X--ray shapes of clusters are less 
flattened spherical for low values of $\Omega_{0}$ than in an Einstein--de 
Sitter Universe. 
Wilson, Cole \& Frenk (1996) also concluded that clusters in an $\Omega_{0} = 
0.2$ Universe are more nearly spherical and centrally concentrated than 
clusters in an $\Omega_{0}=1.0$ Universe. 
On the basis of this result they constructed a new lensing statistic that is 
very sensitive to the value of $\Omega_{0}$ and almost independent of the 
value of the cosmological constant $\Lambda$. 


\subsubsection{Distribution of intrinsic shape}
\label{elldistr}

To investigate which cosmological parameters influence the shape of a 
cluster, we determine the shape of the clusters for all scenarios. 
The present simulations are unconstrained, unlike the simulations of van 
Kampen \& Katgert (1997), and we cannot compare {\it individual} clusters in 
different scenarios. 
Instead, we have to compare the {\it entire distribution} of cluster shapes. 
We describe the shape by two ellipticities which are obtained using the tensor 
of inertia, which was shown to yield reliable cluster ellipticities (de Theije 
{\it et al.} 1995). 
The tensor of inertia is defined as
\begin{equation}
I_{ij} = \sum_{k} x_{ik}x_{jk} / r_{k}^{2} \hspace{1cm} (i,j=1,3) ,
\label{tensormethod}
\end{equation}
where the sum is over all particles in the cluster, $x_{ik}$ $(i=1,3)$ are the 
coordinates of particle $k$ with respect to the cluster centre and $r_{k}$ is 
the distance of that particle to the cluster centre. 
The intrinsic cluster ellipticities are then given by 
\begin{equation}
\epsilon_{1} = 1-\frac{c}{a}, \hspace{1cm} \epsilon_{2} = 1-\frac{b}{a}, 
\label{inteps}
\end{equation}
where $a \ge b \ge c$ are the eigenvalues of the tensor of inertia $I$ (Eq. 
\ref{tensormethod}). 
The triaxiality parameter $T$ was introduced by Franx, Illingworth \& de  
Zeeuw (1990): 
\begin{equation}
T = \frac{\epsilon_{2}(2-\epsilon_{2})}{\epsilon_{1}(2-\epsilon_{1})} = 
\frac{a^{2}-b^{2}}{a^{2}-c^{2}} .
\end{equation}
A value of $T = 1.0$ indicates that a cluster is prolate while a value of 
$T = 0.0$ represents an oblate cluster. 
Values between 0.0 and 1.0 describe triaxial clusters for which $a$, $b$ and 
$c$ all have different values. 

\par Figure \ref{nbodyell} shows the distributions of $\epsilon_{1}$. 
There are no obvious differences among the different scenarios. 
For all scenarios there is a very small number of spherical clusters 
($\epsilon_{1} = 0$) and the largest cluster ellipticity is about 0.8. 
However, the Kolmogorov--Smirnov (KS from here on) test (e.g., Press {\it et 
al.} 1989), shows some differences between different scenarios. 
The $k^{-2}$ scenario with $\sigma_{8}=0.64$ and the HDM scenario are 
significantly different from the SCDM, LCDMb, LCDMc, LCDMd, hCDM and TCDM 
scenarios. 
Clusters in the $k^{-2}$ and HDM scenarios have somewhat smaller ellipticities 
than those in the other scenarios. 
The distributions for different values of $\sigma_{8}$ are always consistent 
with each other for all $\Omega_{0}=1.0$ scenarios (the KS--confidence levels 
for these scenarios are always larger than 0.23). 

\begin{figure}
\hspace*{0.7cm}
\psfig{figure=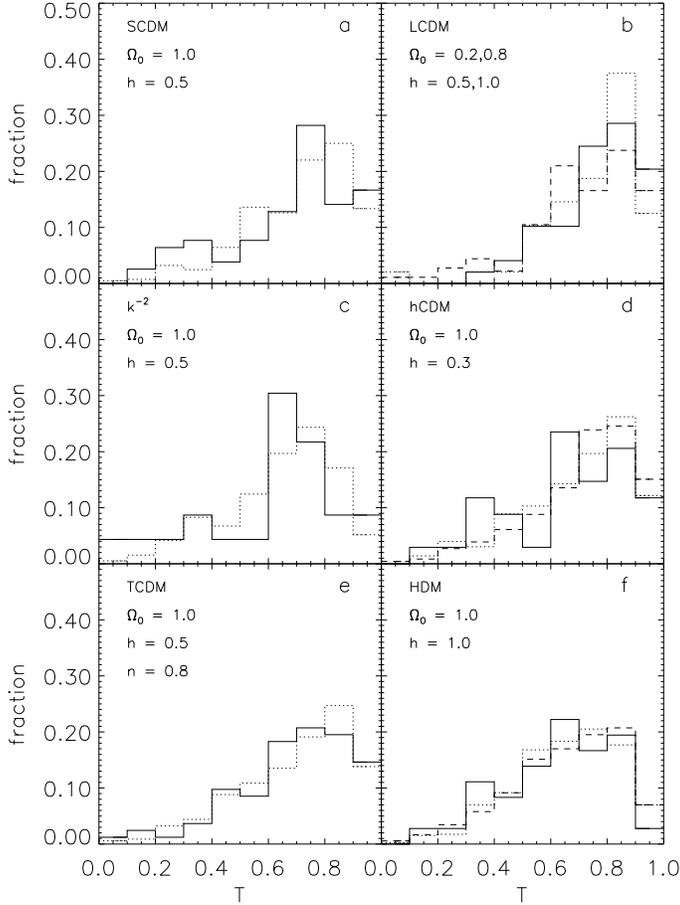,width=7.5cm}
\vspace{0.6cm}
\caption{Distribution of triaxiality parameter $T$ for clusters in the 
different scenarios. 
Only the clusters that have a mass of at least $2.22 \times 10^{14} h^{-1} 
M_{\odot}$ are taken into account. 
A value $T=1.0$ means that the cluster is prolate, $T=0.0$ indicates that the 
cluster is oblate. 
The different panels and curves have the same meaning as in Figure 2.}
\label{nbodyell3d}
\end{figure}

\par The ellipticities within the Abell radius ($r = 1.5 h^{-1}$ Mpc) are 
somewhat larger than those within $r = 1.0 h^{-1}$ Mpc. 
The difference is about $\Delta \epsilon_{1} \approx 0.05$ and occurs in all 
scenarios. 
Qualitatively, such an effect is expected for elongated clusters: the aperture 
bias, as a result of which the cluster seems more spherical than it actually 
is (e.g., de Theije {\it et al.} 1995), is less important for a large aperture 
radius. 
The projected position angle is found to change very little with radius (see 
Section \ref{alignments}). 

\par Figure \ref{nbodyell3d} shows the distribution of the triaxiality 
parameter $T$ for all scenarios. 
The different curves have the same meaning as in Figure \ref{nummass}. 
No obvious differences are found in the distribution of $T$ between 
different scenarios. 
In all scenarios a large number of clusters have $T=0.6-0.9$, indicating that 
most clusters are nearly prolate. 
Perfect oblate clusters ($T=0.0$) are absent in most scenarios. 
The low--$\Omega_{0}$ scenarios contain clusters of slightly larger values of 
$T$. 
The 'strange' distribution of $T$ for the $k^{-2}$ scenario with $\sigma_{8} 
= 0.40$ is probably the result of limited statistics: the number of clusters 
is only 23 (see Table \ref{tababundances}). 

\begin{figure}
\hspace*{1cm}
\psfig{figure=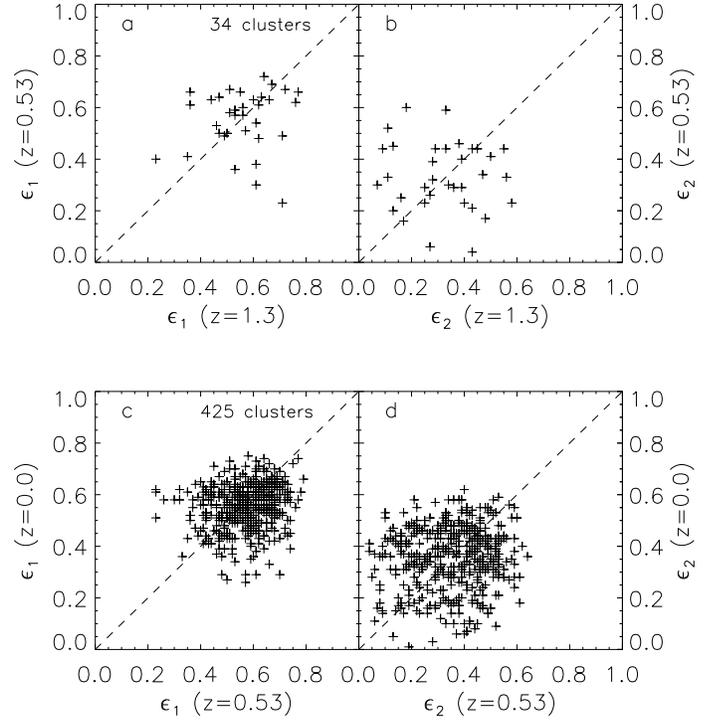,width=7cm}
\caption{Evolution of the shape of clusters in the hCDM scenario. 
Only clusters are used that have a mass of at least $2.22 \times 10^{14} 
h^{-1} M_{\odot}$ at both redshifts. 
{\bf a}: $\epsilon_{1}$ at $z=1.3$ versus $z=0.53$. 
{\bf b}: $\epsilon_{2}$ at $z=1.3$ versus $z=0.53$. 
{\bf c}: $\epsilon_{1}$ at $z=0.53$ versus $z=0$.
{\bf c}: $\epsilon_{2}$ at $z=0.53$ versus $z=0$.}
\label{epsevol1}
\end{figure}

\begin{figure}
\hspace*{1cm}
\psfig{figure=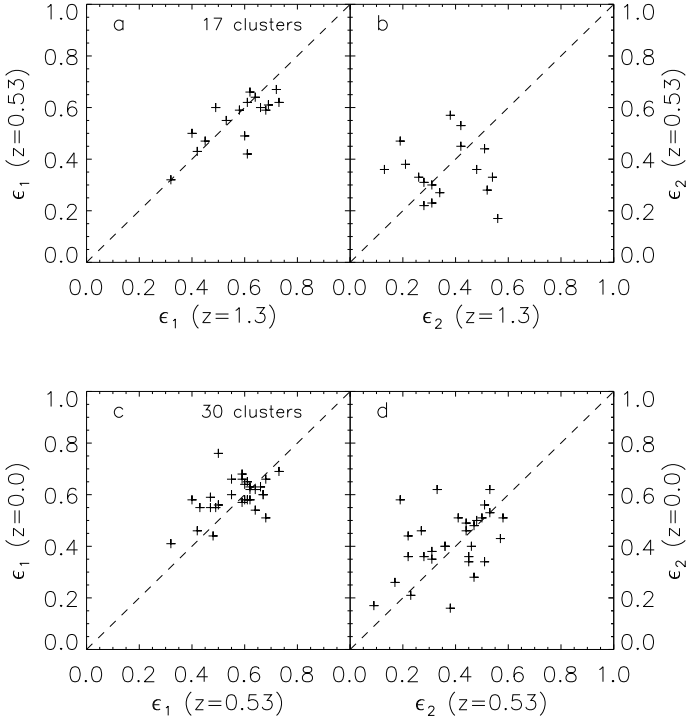,width=7cm}
\caption{Evolution of the shape of clusters in the LCDMb scenario. 
Only clusters are used that have a mass of at least $2.22 \times 10^{14} 
h^{-1} M_{\odot}$ at both redshifts. 
The panels have the same meaning as in Figure 10.}
\label{epsevol0_2}
\end{figure}


\subsubsection{Evolutionary changes in $\epsilon_{1}$, $\epsilon_{2}$ and $T$}

The distributions of $\epsilon_{1}$ for different values of $\sigma_{8}$ in 
the $\Omega_{0}=1.0$ scenarios in Figure \ref{nbodyell} are not significantly 
different. 
This does not necessarily mean that individual clusters have a constant 
$\epsilon_{1}$ and $\epsilon_{2}$. 
To investigate this, we determine $\epsilon_{1}$ and $\epsilon_{2}$ for the 
clusters in the hCDM scenario at different times, as an example for the 
$\Omega_{0}=1.0$ scenarios. 
In Figures \ref{epsevol1}a and b, the values of $\epsilon_{1}$ and 
$\epsilon_{2}$ at $z=1.3$ are compared with those at $z=0.53$. 
For this, only clusters that fulfil the minimum mass limit at both redshifts 
are used. 
We identify the {\it same} clusters at both redshifts by requiring that at 
least half of the particles of the cluster at one redshift is member of the 
cluster at the other redshift, and vice versa. 
This uniquely links clusters at different redshifts. 
The same procedure is repeated for clusters at $z=0.53$ and $z=0.0$ (Figures 
\ref{epsevol1}c and d). 
Although the distributions of $\epsilon_{1}$ and $\epsilon_{2}$ are roughly 
constant in time, individual clusters may show large changes. 
These occur both ways, i.e. $\epsilon_{1}$ and $\epsilon_{2}$ can both 
increase and decrease. 
The mean change in $\epsilon_{1}$ between redshifts $z=1.3$ and 0.53 is 0.00, 
and the {\it r.m.s.} scatter around this line is 0.16. 
Between redshifts $z=0.53$ and the present, the mean change is --0.01 and 
the {\it r.m.s.} scatter is 0.12. 
For $\epsilon_{2}$ these values are --0.01 and 0.20 between $z=1.3$ and 0.53 
and 0.00 and 0.25 between $z=0.53$ and $z=0.00$, respectively. 

\par For the $\Omega_{0}=0.2$ scenarios, this picture does not change 
qualitatively, but quantitatively the changes are somewhat smaller. 
In Figure \ref{epsevol0_2} the same plots are shown as in Figure 
\ref{epsevol1} but now for clusters in the LCDMb scenario. 
Low-$\Omega_{0}$ clusters change their ellipticities in the course of time as 
well, but the dispersions are smaller than in Figure \ref{epsevol1}. 
The mean and {\it r.m.s.} change between $z=1.3$ and 0.53 are 0.02 and 0.08 
for $\epsilon_{1}$ and 0.01 and 0.17 for $\epsilon_{2}$, respectively. 
Between $z=0.53$ and 0.0 these values are --0.03 and 0.09 for $\epsilon_{1}$ 
and --0.03 and 0.24 for $\epsilon_{2}$, respectively. 
However, one should keep in mind that the number of clusters in the LCDMb 
scenario is (much) smaller than in the hCDM scenario, especially at later 
times. 
The results for the LCDMc and LCDMd scenario are similar to those for the 
LCDMb scenario. 

\par Combining the results for all scenarios, we conclude that the mean 
values of $\epsilon_{1,z1}-\epsilon_{1,z2}$, with $z1 \ne z2$, are always 
consistent with zero. 
I.e., there is no significant change in the average cluster ellipticities over 
the time interval studied. 
The {\it r.m.s.} values range from about 0.10 (for $\epsilon_{1}$ and 
$\Omega_{0}<1.0$) to about 0.25 (for $\epsilon_{2}$ and $\Omega_{0}=1.0$). 

\par Using the number--selected cluster samples, the results for the changes 
in $\epsilon_{1}$ and $\epsilon_{2}$ are very similar to those for the 
mass--limited samples. 

\par It is interesting to note that the change in $\epsilon_{2}$ is always 
larger than the change in $\epsilon_{1}$. 
This may be caused by the fact that $\epsilon_{2}$ is always smaller than 
$\epsilon_{1}$ [Eqs. (\ref{inteps})], and therefore its determination is 
somewhat more difficult, as values closer to zero are more difficult to 
measure (de Theije {\it et al.} 1995). 
However, because of the rather large number of particles per cluster, at 
least $100/\Omega_{0}$, this cannot account for the whole effect. 
One possible explanation is the following: a change in $b$ only affects 
$\epsilon_{2}$, whereas a change in $c$ only affects $\epsilon_{1}$. 
If $b$ and $c (\le b)$ are fixed, a change in $a$ has a larger influence on 
$\epsilon_{2}$ than on $\epsilon_{1}$. 
Quantitatively, if one adopts the mean values $\langle \epsilon_{1} \rangle 
\approx 0.55$ and $\langle \epsilon_{2} \rangle \approx 0.35$, one expects on 
the basis of this simple argument that $\Delta \epsilon_{2} = (1-\langle 
\epsilon_{2} \rangle)/(1-\langle \epsilon_{1} \rangle) \Delta \epsilon_{1} 
\approx 1.4 \Delta \epsilon_{1}$. 
This is roughly equal to the actual values of $\Delta \epsilon_{2}/\Delta 
\epsilon_{1}$ that are found, though the scatter in $\Delta \epsilon_{2} / 
\Delta \epsilon_{1}$ between different scenarios is rather large ($\Delta 
\epsilon_{2} / \Delta \epsilon_{1} \approx 1.3-2.0$). 

\par For the hCDM scenario the mean and {\it r.m.s.} changes in $T$ between 
$z=1.3$ and 0.53 are --0.02 and 0.29, respectively. 
Between $z=0.53$ and $z=0.0$ these values are --0.02 and 0.25. 
For the LCDMb scenario, the mean and {\it r.m.s.} changes between $z=1.3$ and 
0.53 are --0.02 and 0.20, respectively, and between $z=0.53$ and $z=0.0$ these 
values are --0.03 and 0.17, respectively. 
The mean value of $T_{z1}-T_{z2}$, with $z1 > z2$, is negative for all 
scenarios, which indicates that, on average, $T$ increases with time. 
That is, clusters become more prolate. 
This may also be the reason why the $\Omega_{0}=0.2$ model clusters have, on 
average, a larger value of $T$. 
They have evolved further than clusters in an $\Omega_{0}=1.0$ scenario. 
The {\it r.m.s.} scatter decreases slightly with time in most scenarios. 

\par Our results, that the values of $\epsilon_{1}$ and $\epsilon_{2}$ change 
in time, while the distribution of $\epsilon_{1}$ is constant in time, may be 
explained as follows: 
at a specific time new particles, that were not within the cluster before, 
enter the cluster. 
For an individual cluster, the new cluster particles may enter the cluster via 
one specific direction and change its shape. 
However, averaged over the cluster population, the new particles fall in 
isotropically and therefore do not change the distribution of ellipticities 
of the cluster population as a whole. 

\par The idea that clusters become more prolate at later times is consistent 
with the findings of Salvador--Sol\'e \& Solanes (1993). 
They concluded that the elongations of clusters are consistent with clusters 
being prolate and that the elongations are mainly produced by the tidal 
interactions of sufficiently massive nearby clusters. 

\begin{figure}
\hspace*{0.7cm}
\psfig{figure=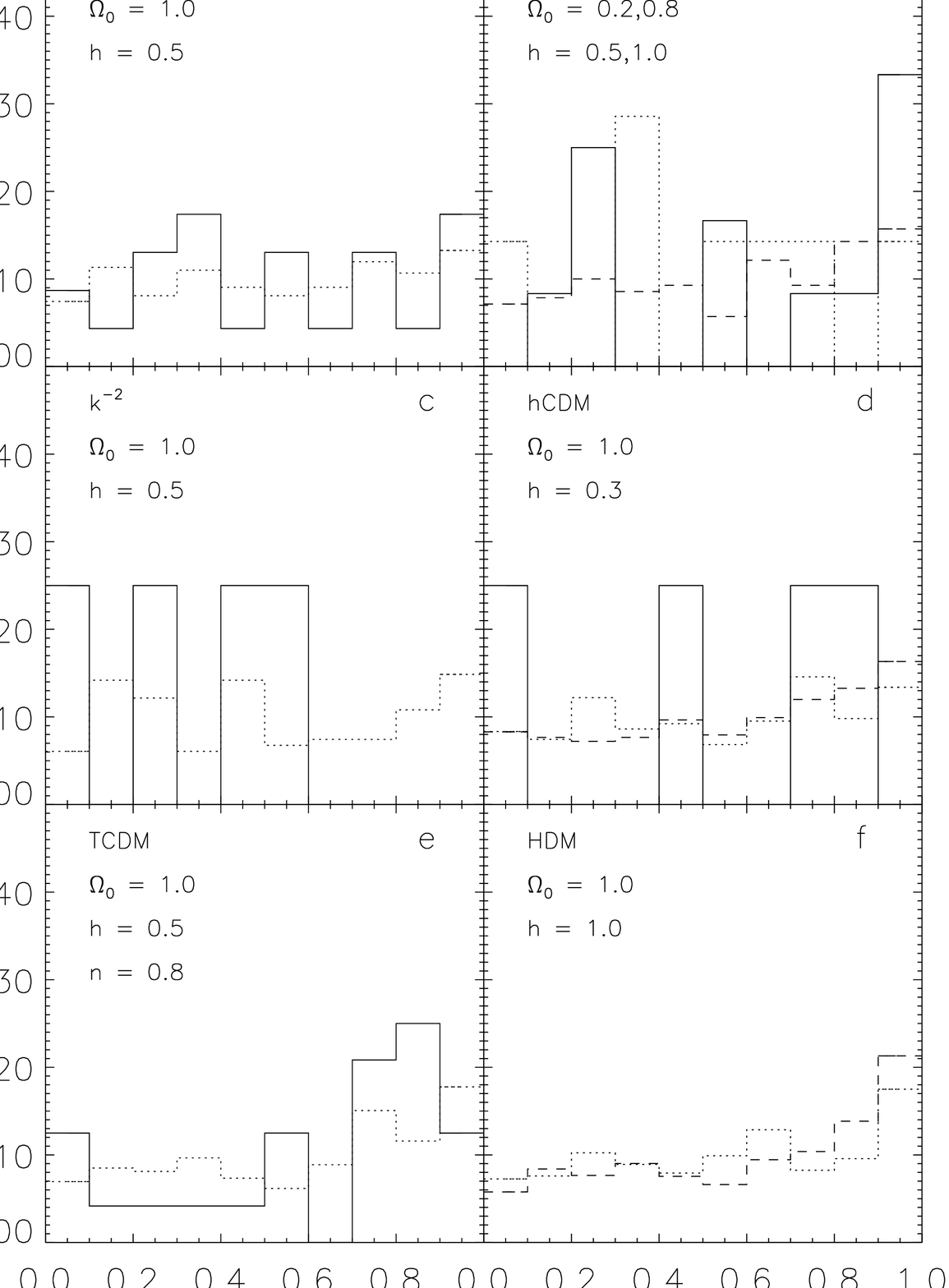,width=7.5cm}
\vspace{0.6cm}
\caption{Angular difference between the cluster major axis and the direction 
to its nearest neighbouring cluster. 
The distributions are for clusters with a mass of at least $2.22 \times 
10^{14} h^{-1} M_{\odot}$, while the nearest neighbour has a mass of at least 
$1.00 \times 10^{14} h^{-1} M_{\odot}$ and is within 20 $h^{-1}$ Mpc. 
The different panels and curves have the same meaning as in Figure 2.}
\label{align1fig}
\end{figure}

\begin{table*}
\centering
\caption{Kolmogorov--Smirnov (KS) confidence levels that the clusters 
in our simulations do {\it not} show an alignment effect. 
The first two columns indicate the scenario and the value of $\sigma_{8}$. 
The quantities with the subscript '144' refer to the number--selected cluster 
samples, whereas all other quantities refer to the mass--limited samples. 
Columns 3 to 5 give the KS--confidence levels that the cluster major axis is 
{\it not} aligned with the direction towards its nearest neighbour. 
The number of clusters involved in the mass--limited samples is given in 
column 3. 
Column 6 to 8 give the KS--confidence levels that the cluster major axis is 
{\it not} aligned with the mass distribution within $10 h^{-1}$ Mpc around it. 
The number of clusters involved in the mass--limited samples is given in 
column 6. 
Column 9 and 10 give the KS--confidence levels that the cluster major axis is 
{\it not} aligned with the major axis of its nearest neighbour cluster. 
The number of clusters involved in the mass--limited samples is given in 
column 3. 
See the text for more details.}
\begin{tabular}{lcrllrllll}
\hline
scenario & $\sigma_{8}$ & $N_{1,3}$ & $P_{\mbox{KS,1}}$ & 
$P_{\mbox{KS,1,144}}$ & $N_{2}$ & $P_{\mbox{KS,2}}$ & $P_{\mbox{KS,2,144}}$ & 
$P_{\mbox{KS,3}}$ & $P_{\mbox{KS,3,144}}$ \\
\hline
SCDM  & 0.46    &  23 & 1.00                  & 1.00                  &  78 & 
 0.13                  & $0.26 \times 10^{-1}$ & 1.00 & 0.70 \\
SCDM  & 0.61    & 309 & 0.65                  & 0.57                  & 398 & 
 $0.79 \times 10^{-6}$ & $0.39 \times 10^{-2}$ & 0.11 & 0.23 \\
      &         &     &                       &                       &     & 
                       &                       &      &      \\
LCDMb & 0.90    &  12 & 0.85                  & $0.26 \times 10^{-1}$ &  49 & 
 $0.54 \times 10^{-1}$ & $0.15 \times 10^{-4}$ & 0.97 & 1.00 \\
LCDMc & 0.90    &   7 & 0.85                  & $0.34 \times 10^{-1}$ &  47 & 
 0.43                  & $0.58 \times 10^{-3}$ & 1.00 & 1.00 \\
LCDMd & 0.90    & 140 & 0.32                  & 0.75                  & 179 & 
 $0.36 \times 10^{-3}$ & $0.92 \times 10^{-3}$ & 0.29 & 0.20 \\
      &         &     &                       &                       &     & 
                       &                       &      &      \\
$k^{-2}$ & 0.46 &   4 & 0.91                  & 1.00                  &  23 & 
 0.55                  & 0.49                  & 0.98 & 0.78 \\
$k^{-2}$ & 0.64 & 148 & 0.97                  & 0.99                  & 191 & 
 $0.30 \times 10^{-3}$ & $0.63 \times 10^{-2}$ & 0.90 & 1.00 \\
      &         &     &                       &                       &     & 
                       &                       &      &      \\
hCDM  & 0.44    &   4 & 1.00                  & 1.00                  &  34 & 
 $0.65 \times 10^{-1}$ & $0.74 \times 10^{-1}$ & 0.70 & 0.61 \\
hCDM  & 0.65    & 336 & 0.26                  & 0.36                  & 425 & 
 $0.24 \times 10^{-7}$ & $0.23 \times 10^{-3}$ & 0.19 & 0.82 \\
hCDM  & 1.00    &1108 & $0.66 \times 10^{-6}$ & $0.86 \times 10^{-2}$ &1168 & 
 $0.24 \times 10^{-18}$ & $0.11 \times 10^{-2}$ & 0.52 & 1.00 \\
      &         &     &                       &                       &     & 
                       &                       &      &      \\
TCDM  & 0.47    &  24 & 0.29                  & 0.66                  &  81 & 
 $0.19 \times 10^{-2}$ & $0.40 \times 10^{-3}$ & 1.00 & 0.60 \\
TCDM  & 0.60    & 259 & $0.93 \times 10^{-2}$ & $0.85 \times 10^{-1}$ & 338 & 
 $0.91 \times 10^{-7}$  & $0.40 \times 10^{-5}$ & 0.20 & 0.94 \\
      &         &     &                       &                       &     & 
                       &                       &      &      \\
HDM   & 0.44    &   0 &                       & 1.00                  &  35 & 
 $0.63 \times 10^{-1}$ & $0.84 \times 10^{-2}$ &      & 0.81 \\
HDM   & 0.66    & 303 & 0.26                  & 0.37                  & 458 & 
 $0.83 \times 10^{-17}$ & $0.33 \times 10^{-6}$ & 0.18 & 0.67 \\
HDM   & 1.00    & 953 & $0.20 \times 10^{-9}$ & $0.11 \times 10^{-3}$ &1063 & 
 $0.91 \times 10^{-47}$ & $0.44 \times 10^{-7}$ & $0.20 \times 10^{-1}$ & 
 0.37 \\
\hline
\hline
\end{tabular}
\label{tabalign}
\end{table*}


\subsection{Alignments}
\label{alignments}

Dekel {\it et al.} (1984) and West {\it et al.} (1989) concluded that the 
relative orientations of cluster major axes with the direction towards 
neighbouring clusters provide a sensitive test for the formation of 
large--scale structure in the Universe. 
However, they used 'only' about 10,000 particles in their simulations, so 
their results may be influenced by resolution effects. 
We study this question again, using our simulations in which we measure 
cluster major axes using the tensor of inertia method (see Section \ref{ell}). 
We consider here only the alignments in 3--D. 
The alignments that result in projected 2--D data will be discussed in Paper 
II. 
Three types of alignment of a cluster with its environment are investigated. 


\subsubsection{Alignment with nearest neighbour}
\label{align1section}

The first type of alignment considered is that between the cluster major axis 
and the direction towards the nearest neighbour cluster. 
This is the sort of alignment that Binggeli (1982) found observationally in 
projection. 
It may arise, e.g., from the tidal force of a neighbouring cluster on the 
cluster under consideration. 
Likewise, the filamentary structure of the mass (or galaxy) distribution may 
induce this type of alignment. 

\par The distribution of the angular difference between the cluster major axis 
and the direction to its nearest neighbour is shown in Figure \ref{align1fig}. 
In this Figure, only clusters with a mass of at least $2.22 \times 10^{14} 
h^{-1} M_{\odot}$ are included, which have a nearest neighbour with a mass of 
at least $1.00 \times 10^{14} h^{-1} M_{\odot}$ and which is closer than 
$20 h^{-1}$ Mpc. 
The results are plotted in terms of $\cos(\Delta\phi)$ because a random 
distribution expresses itself via a uniform distribution of 
$\cos(\Delta\phi)$. 

\par There is some dependence of these distributions on the cosmological 
scenario.  
However, the differences are rather hard to quantify because of the small 
number of cluster pairs involved in some of the scenarios. 
The fourth column of Table \ref{tabalign} gives the confidence levels, derived 
from a KS--test, that the distribution of $\cos(\Delta\phi)$ is consistent 
with a random distribution. 
The third column gives the number of cluster pairs involved in this analysis. 
All scenarios are consistent with a random distribution, except the hCDM, TCDM 
and HDM scenarios with the largest value of $\sigma_{8}$. 
This could be mainly due to the large number of cluster pairs in these 
scenarios. 
If the minimum mass limit for the parent cluster is decreased by a factor of 
two, the two $\Omega_{0}=0.2$ scenarios show a significant alignment effect as 
well, while the $\Omega_{0}=0.8$ scenario shows a marginally significant 
alignment effect. 
Apparently, the significance is mainly restricted by the number of cluster 
pairs in the sample. 
Changing the minimum mass of the nearest neighbour cluster does not influence 
the above results. 
Relaxing the constraint that the nearest neighbour should be within 
$20 h^{-1}$ Mpc of the parent cluster only changes the result for the HDM 
scenario with $\sigma_{8}=0.66$. 
The KS--confidence level for this scenario to have a random distribution of 
$\cos(\Delta\phi)$ then becomes 0.048, a marginally significant alignment. 

\begin{figure}
\hspace*{0.7cm}
\psfig{figure=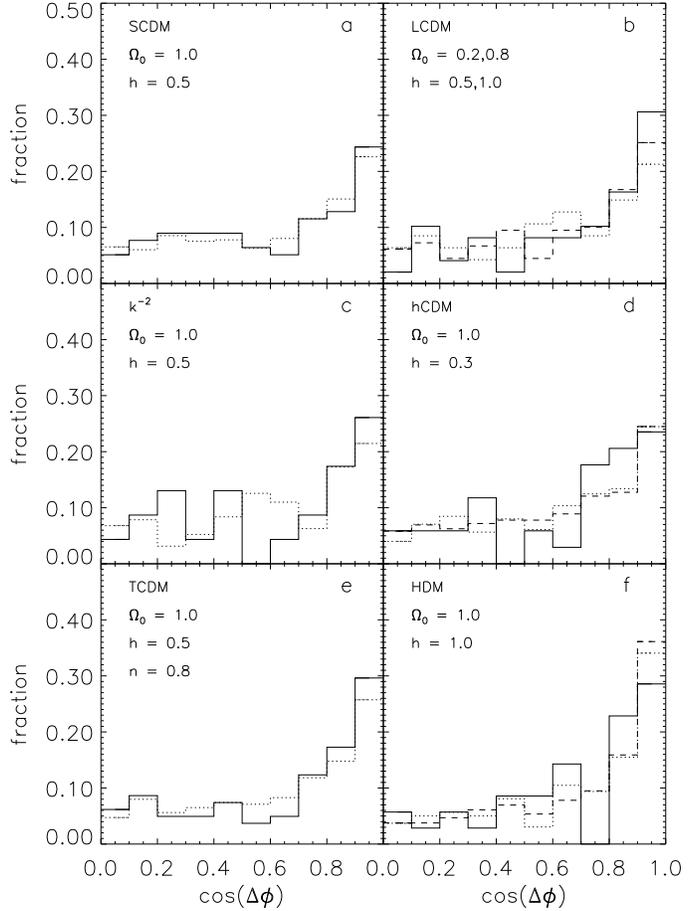,width=7.5cm}
\vspace{0.6cm}
\caption{Angular difference between the cluster major axis and the mass 
distribution within $10 h^{-1}$ Mpc around it. 
The different panels and curves have the same meaning as in Figure 2.}
\label{align2fig}
\end{figure}

\par There is a somewhat stronger cluster alignment if one only considers 
cluster pairs with distances between 10 and $20 h^{-1}$ Mpc. 

\par To check how the above results depend on our definition of the cluster 
sample, the same analysis is done for the number--selected samples. 
The results are given in the fifth column of Table \ref{tabalign}. 
All KS--confidence levels are very similar to those in column 4. 
The significance level is different from that in column 4 only for the 
$\Omega_{0}=0.2$ scenarios, the hCDM and HDM scenarios with $\sigma_{8}=1.00$, 
and the TCDM scenario with $\sigma_{8}=0.60$. 
For the former two scenarios, the alignment effect is just significant for the 
number--selected catalogues while they were not significant for the 
mass--limited catalogues. 
The reason for this difference is the larger number of clusters in the 
number--selected catalogues. 

\par From the table it appears that the significance of the cluster alignment 
with respect to its nearest neighbour increases with $\sigma_{8}$. 
In other words, the alignment effect gets stronger at later times in the 
evolution. 
This can be understood by realizing that it takes some time for the tidal 
torque of a neighbouring cluster to build up this effect. 


\subsubsection{Alignment with environments}
\label{align2section}

The second type of alignment we investigate is that between the cluster major 
axis and the particle distribution around the cluster. 
This is similar to the effect described by Argyres {\it et al.} (1986) and 
Lambas, Groth \& Peebles (1988), who found that galaxy counts are 
systematically high along the line defined by the projected major axis of a 
cluster or of its dominant galaxy. 
The effect extends to at least $15 h^{-1}$ Mpc from the cluster centre. 

\par The distributions of the angular difference between the cluster major 
axis and that of the mass distribution within $10 h^{-1}$ Mpc from the cluster 
are shown in Figure \ref{align2fig}. 
All clusters in the mass--limited cluster catalogues are used for this 
analysis. 
There is a significant alignment in almost all scenarios, showing that a 
cluster is strongly aligned with its surroundings. 
The alignment is usually much stronger than the alignment of the cluster major 
axis with its nearest neighbour, which generally is at a distance smaller than 
$10 h^{-1}$ Mpc. 
The seventh column in Table \ref{tabalign} gives the KS--confidence levels 
that the distribution of $\cos(\Delta\phi)$ results from a random orientation. 
Column 6 gives the number of clusters that is used in this analysis. 
For almost all $\Omega_{0}=1.0$ scenarios, the significance is very high. 
Only the scenarios with the lowest values of $\sigma_{8}$ have lower 
significances. 
The LCDMc scenario does not show a significant alignment effect, while the 
LCDMb scenario shows some effect. 
These results may again be somewhat misleading because these scenarios have 
fewer clusters with masses above the mass limit. 

\par If one uses all particles within $20 h^{-1}$ Mpc of the parent cluster, 
the alignment signal is still present, though sometimes somewhat less 
significant. 
If the surrounding mass distribution in the annulus between 20 and $30h^{-1}$ 
Mpc around the cluster is used, only the HDM scenario with $\sigma_{8}=1.00$ 
shows a positive detection of the alignment. 

\par In column 8 in Table \ref{tabalign} we give the results if the 
number--selected catalogues are used to do the above analysis instead of the 
mass--limited catalogues. 
In this case, almost all scenarios show a significant alignment. 
Apparently, the significance levels for the mass--limited sample are limited 
by the small number of clusters used in some cases. 


\subsubsection{Alignment with nearest neighbour major axis}
\label{align3section}

The third type of alignment that is evaluated is that between the major axes 
of two neighbouring clusters. 
This type of alignment is, of course, highly correlated with the first type 
of alignment that we considered. 
However, it is not exactly the same. 
Consider two clusters which are mutual nearest neighbours. 
If both make an angle of, e.g., 30 degrees with the line connecting them, 
their relative angle is between 0 and 60 degrees. 
Furthermore, being a nearest neighbour is not a commutative property. 
If cluster $j$ is the nearest neighbour of cluster $i$, but has itself cluster 
$k \ne i$ as nearest neighbour, then this type of alignment may differ 
significantly from the first. 

\par As before, only clusters whose mass is larger than $2.22 \times 10^{14} 
h^{-1} M_{\odot}$, with a nearest neighbour with a mass larger than $1.00 
\times 10^{14} h^{-1} M_{\odot}$ and closer than 20 $h^{-1}$ Mpc, are taken 
into account. 
We find that there is no alignment of major axes of neighbouring clusters. 
The ninth column in Table \ref{tabalign} gives the KS--confidence levels for 
the model clusters to show no alignment. 
The number of cluster pairs involved in this analysis is given in column 3. 
Lowering the minimum mass limit of the parent cluster by a factor of two 
does not change this result. 

\par Similarly, the results for this type of alignment do not change if one 
uses the 144 most massive groups in the simulations. 
So the previous results are not due to small number statistics. 
For completeness, the KS--confidence levels for the number--selected sample 
are shown in column 10 of Table 6. 

\begin{table*}
\centering
\caption{Summary of the cosmological parameters that were varied in the 
simulations. 
The first column gives the cosmological parameter. 
The second column shows the parameter values that were used for investigating 
the influence of the parameter. 
The third column gives the identification of the scenarios that are used 
for the investigation. 
The fourth to eighth columns show the other parameters that describe a 
scenario.}
\begin{tabular}{llllccccc}
\hline
parameter & values & scenarios & \multicolumn{6}{c}{other parameter values} \\
\cline{4-9}
          &        & & $\sigma_{8}$ & $\Omega_{0}$ & spectrum & $h$ & $n$ & $\Gamma$ \\
\hline
$\sigma_{8}$ & 0.46--0.61 & SCDM        & & 1.0 & CDM      & 0.5 & 1.0 & 0.5 \\
             & 0.46--0.64 & $k^{-2}$    & & 1.0 & $k^{-2}$ & 0.5 & 1.0 & 0.5 \\
             & 0.44--1.00 & hCDM        & & 1.0 & CDM      & 0.3 & 1.0 & 0.3 \\
             & 0.47--0.60 & TCDM        & & 1.0 & CDM      & 0.5 & 0.8 & 0.5 \\
             & 0.44--1.00 & HDM         & & 1.0 & HDM      & 1.0 & 1.0 & 1.0 \\
             &            &             &       &     &     & & & \\
$\Omega_{0}$ & 0.34--1.00 & SCDM--LCDMd-LCDM034 & $\approx 0.60$ & & CDM & 0.5 & 1.0 & \\
             &            &             &          &     &     & & & \\
spectrum & CDM/$k^{-2}$   & SCDM--$k^{-2}$ & 0.46-0.61 & 1.0 & & 0.5 & 1.0/--2.0 & 0.5 \\
             &            &             &          &     &     & & & \\
$n$          & 1.0/0.8    & SCDM--TCDM  & 0.46-0.61 & 1.0 & CDM & 0.5 & & 0.5 \\
             &            &             &          &     &     & & & \\
$h$          & 0.5/0.3    & SCDM--hCDM  & 0.46-0.61 & 1.0 & CDM & & 1.0 & \\
             & 0.5/1.0    & LCDMb--LCDMc & 0.90 & 0.2 & CDM & & 1.0 & \\
             &            &             &          &     &     & & & \\
$\Gamma$     & 0.20--0.43 & LCDMc       & 0.48-0.90 & 0.2 & CDM & 1.0 & 1.0 & \\
\hline
\hline
\end{tabular}
\label{parcomp}
\end{table*}


\subsubsection{Overall alignment properties}

Summarizing the alignment properties of clusters with their nearest 
neighbouring clusters and their environments, one may conclude that clusters 
tend to be strongly aligned with their surrounding mass distribution in 
almost all scenarios. 
The alignment with the direction towards their nearest neighbour cluster is 
less prominent. 
It is only significant for clusters in the hCDM and HDM scenarios, which have 
the largest value of $\sigma_{8}$, and marginally significant in the LCDMb and 
LCDMc scenarios, both with $\Omega_{0}=0.2$, and the TCDM scenario. 
The alignment between cluster major axes of nearest neighbours is not 
significant in any of the scenarios. 
Even though a cluster may be aligned with the direction towards its nearest 
neighbouring cluster, this nearest neighbour itself is more likely to be 
aligned with its immediate surroundings than with the former cluster. 

\par The different strengths of the various types of alignment suggest that 
although clusters are aligned with their nearest neighbour, they may be 
rotated around this direction with almost random rotation angles. 
Note that clusters are not necessarily each others nearest neighbours. 
For neighbours within 20 $h^{-1}$ Mpc, there is only a 52--60\% probability 
that cluster $j$ has cluster $i$ as its nearest neighbour if $j$ is nearest 
neighbour of $i$. 
This probability is almost identical in all scenarios and causes nearest 
neighbour cluster major axes to be less aligned than the major axis of a 
cluster with respect to the direction towards its nearest neighbour. 


\section{RELATIONS BETWEEN COSMOLOGICAL PARAMETERS AND CLUSTER PROPERTIES}
\label{parameters}

Our set of scenarios is chosen in such a way that one can compare scenarios 
that differ in the value of one cosmological parameter only. 
The influence of this particular parameter on the cluster properties can then 
be investigated. 
One may even try to get scaling relations which describe the dependence of 
specific cluster properties on the values of the cosmological parameters. 
This analysis is complicated by the fact that the clusters themselves can be 
very different in different scenarios. 
For example, the clusters that are identified in the $\Omega_{0}=0.2$ 
scenarios have a much smaller mass than clusters in the $\Omega_{0}=1.0$ 
scenarios (see Section \ref{mass}). 
If one wants to investigate the influence of $\Omega_{0}$ on, e.g., the 
cluster shape, one can either use the whole cluster sample or only the 
clusters in a specific mass range. 
In the former case, the correlation between cluster mass and shape (de Theije 
{\it et al.} 1995) will influence the results. 

\par In Table \ref{parcomp} we summarize the parameters of which the effect on 
cluster properties are investigated. 
Column 2 gives the probed values of the particular parameter. 
Column 3 shows the scenarios used for the comparison. 
As discussed before (Section \ref{simulations}), for the spectral parameter 
$\Gamma$ we can only compare scenarios for which the value of $\sigma_{8}$ 
differs as well. 
So those scenarios will not yield a direct indication of the influence of 
$\Gamma$ only, but of the combination of $\Gamma$ and $\sigma_{8}$. 
In this Section, we will only discuss the combinations of cosmological 
parameters and cluster properties that appear correlated. 


\subsection{R.m.s. mass fluctuation on scales of $8 h^{-1}$ Mpc, 
$\sigma_{8}$}
\label{sigma8}

When one compares all scenarios which differ only by the value of 
$\sigma_{8}$, one finds that a larger value of $\sigma_{8}$ produces a larger 
number of clusters with a mass of at least $2.22 \times 10^{14} h^{-1} 
M_{\odot}$ (see Table \ref{tababundances}), and that each cluster individually 
contains more mass. 
These correlations are consistent with previous findings of, e.g., White, 
Efstathiou \& Frenk (1993) and Eke {\it et al.} (1996). 
Using the functional form $N \propto \sigma_{8}^{c_{1}+c_{2}\sigma_{8}}$ (Eke 
{\it et al.} 1996), one obtains $c_{1} \approx 5.5-6.0$ for all 
$\Omega_{0}=1.0$ scenarios. 
The parameter $c_{2}$ depends on the scenario and ranges from --5.1 for the 
HDM scenario to $1.4$ for the $k^{-2}$ scenario. 
Note that an exponential of the form $N \propto \exp(c\,\sigma_{8})$ does not 
provide a reasonable fit. 

\par For the $\Omega_{0}=1.0$ scenarios, the fraction of 
high--$\sigma_{\mbox{los}}$ clusters increases with $\sigma_{8}$ only for 
the hCDM and HDM scenarios, as does the largest value of 
$\sigma_{\mbox{los}}$. 
The median value of $\sigma_{\mbox{los}}$ is constant if one restricts the 
analysis to clusters in the mass range $1.5 \times 10^{14} h^{-1} M_{\odot} 
\le M \le 2.0 \times 10^{14} h^{-1} M_{\odot}$. 
This is expected because $\sigma_{\mbox{los}}^{2}$ correlates with $M$ (see 
Figure \ref{sigma_m}), and using a fixed and very small mass range will result 
in clusters which have almost identical values for $\sigma_{\mbox{los}}$. 

\par These results are independent of the mass limit that one applies, at 
least in the range $1.0-5.0 \times 10^{14} h^{-1} M_{\odot}$. 
These results cannot be checked directly for $\Omega_{0} < 1.0$ from our 
scenarios. 

\par If one uses the set of the 144 most massive clusters, the median value of 
$\sigma_{\mbox{los}}$ increases with $\sigma_{8}$ for the 
$\mathsf{\Omega_{0}=1.0}$ scenarios and is constant in time for the 
$\mathsf{\Omega_{0}<1.0}$ scenarios. 
In particular, for the SCDM scenario one gets $\sigma_{\mbox{los,median}} 
\propto \sigma_{8}^{0.90 \pm 0.01}$, for the hCDM scenario 
$\sigma_{\mbox{los,median}} \propto \sigma_{8}^{\approx 0.84 \pm 0.01}$, and 
for the HDM scenario $\sigma_{\mbox{los,median}} \propto \sigma_{8}^{\approx 
0.78 \pm 0.01}$. 
The indicated errors are due to the fitting only. 

\par The particle mass of the clusters increase with $\sigma_{8}$ as well. 
The fractional increase is about twice as large as for 
$\sigma_{\mbox{los,median}}$, as is expected from the virial theorem. 
Although this is only shown for the $\Omega_{0}=1.0$ scenarios, White {\it et 
al.} (1993) showed that it holds for open scenarios as well. 
The median value of the cluster particle mass seems to increase significantly 
with $\sigma_{8}$ only for the hCDM and HDM scenarios. 
For the hCDM scenario the dependence is given by $M_{\mbox{median}} \propto 
\sigma_{8}^{0.31 \pm 1.58}$ while for the HDM scenario the best--fitting 
relation is $M_{\mbox{median}} \propto \sigma_{8}^{0.33 \pm 0.76}$. 
The errors in the exponents are so large that the correlations are yet 
insignificant. 
For all other scenarios the median value of the cluster particle mass is 
nearly constant. 
These results are independent of the mass limit that one applies, at least in 
the range $1.0-5.0 \times 10^{14} h^{-1} M_{\odot}$. 
If the number--selected cluster set is used, the results are similar to those 
for the median value of the cluster velocity dispersion. 
For the SCDM scenario $M_{\mbox{median}} \propto \sigma_{8}^{1.63 \pm 0.46}$, 
for the hCDM scenario $M_{\mbox{median}} \propto \sigma_{8}^{1.48 \pm 0.61}$, 
and for the HDM scenario $M_{\mbox{median}} \propto \sigma_{8}^{1.41 \pm 
1.01}$. 
For the low--$\Omega_{0}$ scenarios the median value of the cluster particle 
mass is constant in time. 

\par A larger $\sigma_{8}$ produces larger cluster peculiar velocities (see 
Table \ref{tabpecvel}). 
The dispersion of the best--fitting Gaussian distribution to the 1--D cluster 
peculiar velocity distribution scales, on average, linearly with $\sigma_{8}$, 
with a scatter of about 10\%. 
This linear relation is identical to that predicted by linear perturbation 
theory ($v_{\mbox{pec}} \propto \Omega^{0.6} \sigma_{8}$; e.g., Peebles 1993). 

\par The normalization $\sigma_{8}$ has a large effect on the alignment 
properties of clusters. 
Especially the alignment of a cluster with its nearest neighbour and with the 
surrounding mass distribution are influenced by $\sigma_{8}$ (see Table 
\ref{tabalign}). 
Scenarios with a larger value of $\sigma_{8}$ contain clusters that are 
better aligned both with the direction towards their nearest neighbour and 
with their surroundings. 


\subsection{Density parameter $\Omega_{0}$}
\label{omega}

Comparing the SCDM scenario with $\sigma_{8}=0.61$ with the LCDMd scenario, it 
is clear that a higher value of the density parameter $\Omega_{0}$ will 
produce more clusters with a mass of at least $2.22 \times 10^{14} h^{-1} 
M_{\odot}$. 
To quantify this, we also include the LCDMb scenario at the epoch where 
$\sigma_{8} \approx 0.6$. 
This is between $a_{\mbox{exp}}=0.44$ and $a_{\mbox{exp}}=0.55$, and the value 
of $\Omega$ is then equal to 0.34. 
We will refer to this scenario as LCDM034 from now on. 
The number of clusters with a mass of at least $2.22 \times 10^{14} h^{-1} 
M_{\odot}$ is about 23 in this scenario. 
Using the functional form $N \propto \Omega^{c_{1}+c_{2}\,\Omega}$ to describe 
the number of clusters with a mass of at least $2.22 \times 10^{14} h^{-1} 
M_{\odot}$, analogous to Eke {\it et al.} (1996), one obtains $c_{1} = 1.95$ 
and $c_{2} = 2.09$. 
An exponential of the form $N \propto \exp(c\,\Omega_{0})$ does not provide a 
good fit. 

\par
\begin{figure*}
\centering
\psfig{figure=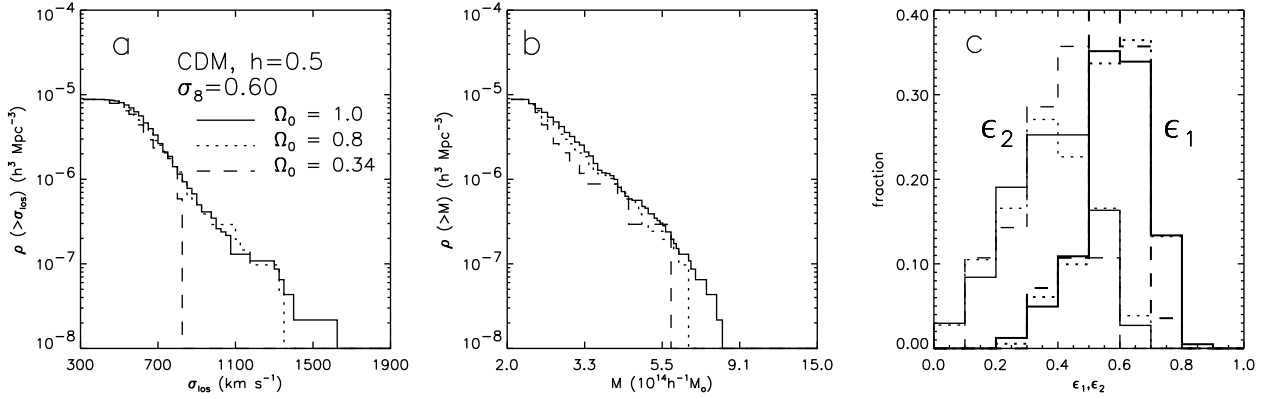,width=17cm}
\caption{{\bf a}: Cumulative distribution of cluster velocity dispersion 
for the SCDM scenario with $\sigma_{8}=0.61$ ($\Omega_{0}=1.0$, solid line), 
the LCDMd scenario ($\Omega_{0}=0.8$, dotted line) and the LCDM034 scenario 
($\Omega_{0}=0.34$, dashed line). 
Only clusters in the mass--limited samples are included and all three 
distributions have been scaled to a total cluster density of $8.6 \times 
10^{-6} h^{3}$ Mpc$^{-3}$. 
{\bf b}: Same as {\bf a}, but for cluster masses. 
{\bf c}: Distributions of cluster ellipticities. 
The thick lines are the distributions of $\epsilon_{1}$, whereas the 
thin lines indicate the distributions of $\epsilon_{2}$. 
Only clusters with a mass of at least $2.22 \times 10^{14} h^{-1} M_{\odot}$
are used. 
The curves have the same meaning as in panel {\bf a}.} 
\label{diffOmega}
\end{figure*}

 The density parameter influences the velocity dispersion of clusters 
above a certain mass. 
This can be seen in Figure \ref{diffOmega}a. 
The difference in $\rho(>\sigma_{\mbox{los}})$ between the SCDM and LCDMd 
scenarios is not very significant, but that between the SCDM and LCDM034 
scenarios is. 
The latter lacks the high--$\sigma$ clusters that the SCDM scenario contains. 
If one uses only the 144 most massive groups in both scenarios one finds that 
the velocity dispersions of these are larger for higher $\Omega_{0}$, 
$\sigma_{\mbox{los,median}} \propto \Omega_{0}^{\approx 0.5}$. 

\par Figure \ref{diffOmega}b shows the cumulative distribution of cluster 
mass for the SCDM, LCDMd and LCDM034 scenarios for all clusters in the 
mass--limited catalogues.
All distributions are scaled to the same number of clusters. 
The SCDM scenario has a few clusters that have a somewhat larger mass than the 
most massive clusters in the LCDMd scenario. 
The curve for the LCDMd scenario is systematically, though only slightly, 
below that of the SCDM scenario. 
This is consistent with the findings of, e.g., White {\it et al.} (1993). 
The same conclusions hold if one compares the SCDM scenario with the LCDM034 
scenario, but the differences are then even larger. 
If one uses the number--selected catalogues in both scenarios, the cluster 
mass is largest for the $\Omega_{0}=1.0$ scenario (see Figure \ref{nummass}). 
The median value of the cluster particle mass for clusters in the 
number--selected sample scales as $M_{\mbox{median}} \propto 
\Omega_{0}^{\approx 0.8}$. 

\par In Figure \ref{diffOmega}c we show the distributions over 3--D 
ellipticities for the SCDM, LCDMd and LCDM034 scenarios and for clusters above 
the mass limit of $2.22 \times 10^{14} h^{-1} M_{\odot}$. 
No significant differences between these distributions are detected. 
This may seem surprising because previous studies showed that the cluster 
ellipticity {\it increases} with $\Omega_{0}$ (Mohr {\it et al.} 1995, de 
Theije {\it et al.} 1995). 
However, the mass range involved here differs substantially between the 
different scenarios (see Section \ref{cluster_definition}). 
As was pointed out by Struble \& Ftaclas (1994) and de Theije {\it et al.} 
(1995), the more massive clusters are less elongated than the less massive 
ones. 
This anti--correlation between $\epsilon$ and $M$ is also detected in the 
present simulations. 

\begin{figure}
\centering
\psfig{figure=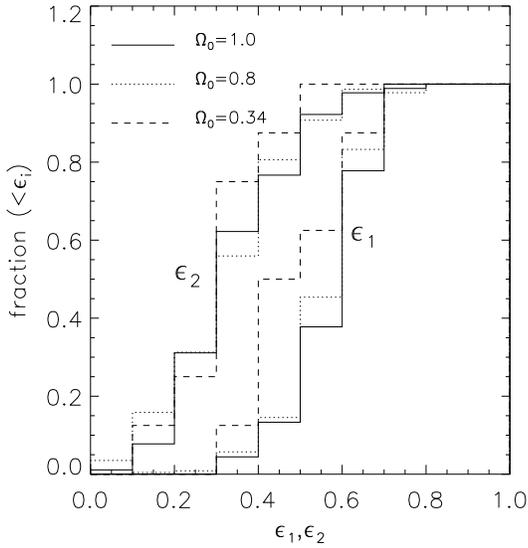,width=8cm}
\caption{Cumulative distributions of cluster ellipticity $\epsilon_{1}$ 
(thick lines) and $\epsilon_{2}$ (thin lines) for the SCDM scenario (solid 
lines, 90 clusters), the LCDMd scenario (dotted lines, 227 clusters) and the 
LCDM034 model (dashed lines, 8 clusters). 
Only clusters that have a mass in the range $(1.5-2.0) \times 10^{14} h^{-1} 
M_{\odot}$ are used. 
The mean values of $\epsilon_{1}$ are 0.62, 0.60 and 0.53 for the SCDM, LCDMd 
and LCDM034 scenarios, respectively.}
\label{epsdiffomega}
\end{figure}

\par To investigate whether $\epsilon_{1}$ and $\epsilon_{2}$ are smaller for 
clusters in low--$\Omega_{0}$ scenarios when clusters of the same mass are 
considered, we show in Figure \ref{epsdiffomega} the cumulative distributions 
of $\epsilon_{1}$ (thick lines) and $\epsilon_{2}$ (thin lines) for the SCDM 
scenario with $\sigma_{8}=0.61$, the LCDMd scenario (having $\sigma_{8}=0.60$) 
and the LCDM034 scenario (having $\sigma_{8} \approx 0.60$). 
These scenarios differ only in the value of $\Omega_{0}$. 
Only clusters in the mass range $1.5 \times 10^{14} h^{-1} M_{\odot} \le M \le 
2.0 \times 10^{14} h^{-1} M_{\odot}$ are included. 
This is to eliminate the correlation between $\epsilon_{i}$ and $M$. 
It is clear that clusters in the same mass range are less flattened in the 
$\Omega_{0}=0.8$ scenario than in the $\Omega_{0}=1.0$ scenario, consistent 
with the previous findings of Mohr {\it et al.} (1995) and de Theije {\it et 
al.} (1995). 
The mean values of $\epsilon_{1}$ are 0.62, 0.60 and 0.53 for the SCDM, LCDMd 
and LCDM034 scenarios, respectively, with dispersions of 0.10, 0.11 and 0.13. 
The mean values of $\epsilon_{2}$ are 0.39, 0.38 and 0.34, respectively, with 
dispersions of 0.15, 0.16 and 0.11. 
Interpolating linearly, we find $\langle \epsilon_{1} \rangle = 0.48 + 0.13\,
\Omega_{0}$ and $\langle \epsilon_{2} \rangle = 0.31 + 0.08\,\Omega_{0}$ for 
clusters in the same mass range. 

\par The dispersions of the fitted Gaussian distributions to the distributions 
of cluster peculiar velocity indicate that a somewhat smaller value of 
$\Omega_{0}$ results in slightly smaller cluster peculiar velocities, 
consistent with previous work of, e.g., Bahcall {\it et al.} (1994) and 
Gramann {\it et al.} (1995). 
The difference in dispersions between the SCDM and LCDMd scenarios is 28 km 
s$^{-1}$, or about 13\%, while the difference between the SCDM and LCDM034 
scenarios is 42 km s$^{-1}$, or about 17\%. 
As linear perturbation theory predicts a relation $v_{\mbox{pec}} \propto 
\Omega^{0.6}$ for constant $\sigma_{8}$ (e.g., Peebles 1993), we fit the 
power--law relation $\sigma_{\mbox{pec,1D}} \propto \Omega^{\gamma}$. 
This does not provide a good fit. 
Neither does an exponential $\sigma_{\mbox{pec,1D}} \propto 
\exp(b\,\Omega_{0})$ give an acceptable fit. 

\begin{figure*}
\centering
\psfig{figure=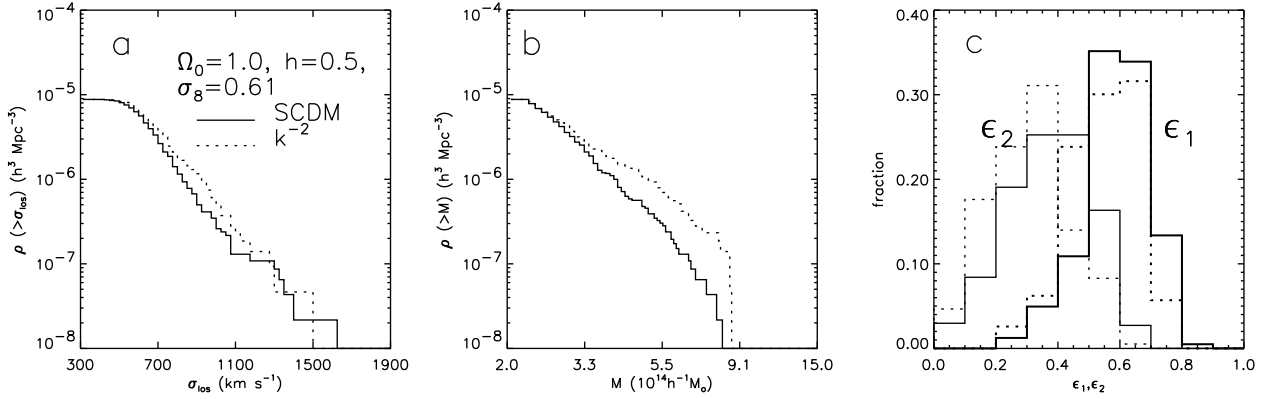,width=17cm}
\caption{Same as Figure \ref{diffOmega}.
The solid line is for the SCDM scenario with $\sigma_{8}=0.61$ and the dotted 
line is for the $k^{-2}$ scenario with $\sigma_{8}=0.64$.}
\label{diffspectrum}
\end{figure*}


\subsection{Spectrum}
\label{spectrum}

Comparing the SCDM with the $k^{-2}$ scenario, both with $\sigma_{8}=0.46$ 
or with $\sigma_{8} \approx 0.61-0.64$, one finds that the number density of 
clusters is about twice as large for the SCDM scenario as for the $k^{-2}$ 
scenario. 
The SCDM spectrum has more power on scales less than $8 h^{-1}$ Mpc, as its 
effective power law index in this $k$--region is --1. 
Apparently, this extra power on the somewhat smaller ($\approx 2 h^{-1}$ Mpc) 
scales stimulates cluster formation considerably through merging of smaller 
clumps. 

\par The comparison between the SCDM and $k^{-2}$ scenarios with $\sigma_{8} 
= 0.61-0.64$ is shown in Figure \ref{diffspectrum}. 
The line--of--sight velocity dispersions are slightly larger for the $k^{-2}$ 
scenario. 
The differences in cluster mass between both scenarios are larger. 
Clusters in the $k^{-2}$ scenario have larger masses than those in the SCDM 
scenario. 

\par The shape of the power spectrum has some influence on the distribution 
over ellipticities. 
The ellipticities for the SCDM spectrum are somewhat larger than those for the 
$k^{-2}$--spectrum, especially for $\sigma_{8} \approx 0.61-0.64$. 
The KS--confidence levels for both distributions to be the same are 0.19 (for 
$\epsilon_{1}$) and 0.24 (for $\epsilon_{2}$) for $\sigma_{8}=0.46$, and 
$3.6 \times 10^{-3}$ (for $\epsilon_{1}$) and $1.1 \times 10^{-5}$ (for 
$\epsilon_{2}$) for $\sigma_{8} \approx 0.61-0.64$. 
The latter are thus significant. 
If one uses the 144 most massive groups in both scenarios, the 
KS--confidence levels are very similar to those for the mass--limited samples. 

\par The shape of the power spectrum has a significant effect on the cluster 
peculiar velocity. 
For the $k^{-2}$ scenario, the cluster peculiar velocity is almost twice as 
large as for the SCDM scenario. 
This is due to the extra power of the $k^{-2}$ scenario on the very large 
scales. 
The results for the number--selected samples are the same. 

\par The index of the primordial spectrum $n$ only has some influence on the 
cluster number densities and peculiar velocities. 
For the number densities, the number of clusters scales as $N \propto 
n^{\delta}$, with $\delta \approx 1.0$, although for various values of 
$\sigma_{8}$ the value of $\delta$ varies from $0.8$ to $1.2$. 
The peculiar velocity is larger by about 20 km s$^{-1}$ (or about 8\%) for the 
TCDM scenario with $n=0.8$, due to the extra large--scale power. 
Using the functional form $\sigma_{\mbox{pec,3D}} \propto n^{-\tau}$ one gets 
$\tau = 0.41 \pm 0.20$. 
Because $n$ does not correlate strongly with the cluster properties, its value 
can be chosen in order to fit the COBE data. 
Cen {\it et al.} (1992) found that $n = 0.7-0.8$ is the most interesting 
range. 


\subsection{Hubble parameter $h$}
\label{h}

For the SCDM and hCDM scenarios with $\sigma_{8} \approx 0.61-0.65$ and for 
the LCDMb and LCDMc scenarios, the number of clusters with a mass of at least 
$2.22 \times 10^{14} h^{-1} M_{\odot}$ is almost identical, so $h$ does not 
have a large influence on the cluster number densities. 
Of course, there would be a large difference in the number of clusters if 
one would express the mass of a cluster in units of $M_{\odot}$ instead of 
$h^{-1} M_{\odot}$. 
In that case, the LCDMb scenario would have 201 clusters with $M > 2.22 \times 
10^{14} M_{\odot}$. 
Qualitatively, this is expected as the Universe is twice as old for 
the LCDMb scenario as it is for the LCDMc scenario (see the last column of 
Table \ref{models}). 
Using only the comparison between the LCDMb and LCDMc scenario, one finds that 
the number of clusters with a mass of at least $2.22 \times 10^{14} M_{\odot}$ 
scales as $N \propto h^{-\alpha}$, with $\alpha$ increasing monotonically 
from 0.49 at $a_{\mbox{exp}} = 0.55$ to 2.29 at $a_{\mbox{exp}} = 1.00$. 

\par A lower value of $h$ results in slightly larger cluster peculiar 
velocities (see Table \ref{tabpecvel}). 
This is especially true if $\sigma_{8}$ is not too high, i.e., $\sigma_{8} \le 
0.6$. 
For the dispersion of the 3--D peculiar velocity distribution we find the 
empirical relation $\sigma_{\mbox{pec,3D}} \propto h^{-\beta}$, where $\beta 
\approx 0.25-0.35$. 

\par The influence of $h$ on the cluster properties is similar if one uses the 
144 most massive groups. 


\subsection{Spectral parameter $\Gamma$}
\label{gamma}

The number of clusters with mass larger than $2.22 \times 10^{14} h^{-1} 
M_{\odot}$ decreases with increasing $\Gamma$. 
For the LCDMc scenario one gets $N \propto \Gamma^{-1.55 \pm 0.02}$, though 
one has to keep in mind that the value of $\sigma_{8}$ changes as well, from 
0.47 to 0.90. 
The actual relation between $N$ and $\Gamma$ may thus be even steeper. 
A similar result holds for the LCDMb scenario, $N \propto \Gamma^{-1.43 \pm 
0.01}$. 

\par The median values of the cluster particle mass and velocity dispersion 
hardly change with $\Gamma$ for the LCDMb and LCDMc models. 

\par The peculiar velocities increase with $\Gamma$ according to $\Gamma^{0.32 
\pm 0.01}$ for the LCDMc scenario, and according to $\Gamma^{0.38 \pm 0.01}$ 
for the LCDMb scenario. 
This is surprising, as a larger $\Gamma$ means that there is less large--scale 
power (Efstathiou {\it et al.} 1992). 
Furthermore, for the epochs where $\Gamma$ is smaller, $\sigma_{8}$ is larger. 
This would suggest an even steeper increase of the peculiar velocity towards 
the epochs when $\Gamma$ is small. 
Possibly the results are influenced by the rather small number of clusters. 


\section{CONCLUSIONS AND DISCUSSION}
\label{conclusions}

The aim of this paper was fourfold: 
(1) Present the set of simulations which will be used in Paper II to select 
the scenario that is most consistent with the observations. 
(2) Study the intrinsic properties of clusters of galaxies for different 
cosmological scenarios. 
(3) Investigate which cosmological parameters have the largest influence on 
these cluster properties. 
(4) Obtain scaling relations between the cosmological parameters and the 
cluster properties. 
These scaling relations can be used to 'interpolate' between existing 
scenarios.

\par The following conclusions may be drawn from the present analysis, for 
the range of parameters studied and the scenarios studied: 

\begin{itemize}

\item $\sigma_{8}$ (in the range 0.44--1.00): 

\begin{itemize}

\item The normalization $\sigma_{8}$ correlates positively with the cluster 
number density, as is expected. 
This is consistent with the earlier findings of, e.g., White {\it et al.} 
(1993). 
Fitting $N \propto \sigma_{8}^{c_{1}+c_{2}\,\sigma_{8}}$, we get $c_{1} 
\approx 5.5-6.0$ for all $\Omega_{0}=1.0$ scenarios, while $c_{2}$ differs 
between various scenarios. 

\item The median value of the cluster line--of--sight velocity dispersions is 
almost independent of $\sigma_{8}$ for the $\Omega_{0}=1.0$ scenarios. 
Only for the hCDM and HDM scenarios there is a slight correlation, although 
the errors are large. 
These results are independent of the mass limit that one applies, at least in 
the range $1.0-5.0 \times 10^{14} h^{-1} M_{\odot}$. 
Using the number--selected cluster samples, $\sigma_{\mbox{los,median}} 
\propto \sigma_{8}^{0.78-0.90}$ for the $\Omega_{0}=1.0$ scenarios. 

\item The median value of the cluster particle mass increases somewhat with 
$\sigma_{8}$ for the hCDM and HDM scenarios, though the errors are very large. 
These results are independent of the mass limit that one applies, at least in 
the range $1.0-5.0 \times 10^{14} h^{-1} M_{\odot}$. 
Using the number--selected cluster samples, $M_{\mbox{median}} \propto 
\sigma_{8}^{1.41-1.63}$. 

\item The value of $\sigma_{8}$ does not influence the cluster ellipticities. 
Although for individual clusters the ellipticity changes, the {\it 
distribution} of ellipticity is independent of $\sigma_{8}$. 

\item A larger value of $\sigma_{8}$ results in significantly larger cluster 
peculiar velocities. 
Empirically, we found $\sigma_{\mbox{pec,1D}} \propto \sigma_{8}$, as is 
predicted by linear theory. 

\item $\sigma_{8}$ does affect the cluster alignment with its nearest 
neighbouring cluster and with its surroundings. 
The angular difference between the major axes of neighbouring clusters does 
not, or only marginally, depend on $\sigma_{8}$. 

\end{itemize}

\item $\Omega_{0}$ (in the range 0.34--1.00): 

\begin{itemize}

\item A larger value of $\Omega_{0}$ results in an increase in the number of 
clusters. 
Fitting the functional form $N \propto \Omega_{0}^{c_{1}+c_{2}\,\Omega_{0}}$ 
we obtained $c_{1} = 1.95$ and $c_{2} = 2.09$. 
Changing $\Omega_{0}$ changes the shape of the distribution of cluster 
line--of--sight velocity dispersions for clusters above a certain mass, 
and it results in considerably larger velocity dispersions for the 144 most 
massive clusters, $\sigma_{\mbox{los,median}} \propto 
\Omega_{0}^{\approx 0.39 \pm 0.08}$. 

\item A larger $\Omega_{0}$ produces more high--mass clusters both in absolute 
and in relative sense, consistent with earlier results of White {\it et al.} 
(1993) and Jing \& Fang (1994). 
This holds for both the mass--limited and the number--selected cluster 
catalogues. 
For the latter catalogues, $M_{\mbox{median}} \propto \Omega_{0}^{\approx 
0.97 \pm 0.20}$. 

\item A low value of $\Omega_{0}$ produces more spherical clusters for a 
specific mass range. 
This was already concluded by de Theije {\it et al.} (1995) and Mohr {\it et 
al.} (1995). 
However, for the sample of clusters above a minimum mass threshold or for the 
sample of the $N$ most massive clusters, the correlation between $\Omega_{0}$ 
and $\epsilon$ disappears. 
This is because the cluster mass is very different in the different scenarios, 
and because $\epsilon$ and $M$ are anti--correlated. 

\item A larger value of $\Omega_{0}$ results in an increase in the cluster 
peculiar velocity, consistent with earlier findings of, e.g., Bahcall {\it 
et al.} (1994). 
However, a fit of the form $\sigma_{\mbox{pec,1D}} \propto 
\Omega_{0}^{\gamma}$ as is predicted by linear theory, where $\gamma = 0.6$, 
does not provide a good representation. 

\end{itemize}

\item Spectrum (SCDM versus $k^{-2}$):

\begin{itemize}

\item The shape of the spectrum clearly influences the cluster number density. 
The SCDM spectrum produces more clusters than does the $k^{-2}$ spectrum. 
The clusters in both scenarios have a similar velocity dispersion. 

\item The spectrum has some influence on cluster mass as well. 
Clusters in the $k^{-2}$ scenario have a larger mass than in the SCDM 
scenario. 

\item The SCDM spectrum produces clusters that are somewhat more elongated 
than does the $k^{-2}$ power spectrum. 
The difference is only significant for $\sigma_{8} \ge 0.6$. 

\item The $k^{-2}$ spectrum produces cluster peculiar velocities that are 
almost twice as large as for the SCDM spectrum. 
This is due to the extra power on large scales. 

\item For smaller values of $n$, the cluster peculiar velocity is larger 
because of the extra power on large scales. 

\end{itemize}

\item $h$ (in the range 0.5--1.0):

\begin{itemize}

\item The Hubble--parameter $h$ affects the cluster peculiar velocity 
slightly: the 3--D cluster peculiar velocities scale as $h^{-\beta}$, with 
$\beta \approx 0.25-0.35$. 
This is because for smaller values of $h$ the Universe is older. 

\end{itemize}

\item $\Gamma$ (in the range 0.20--0.43):

\begin{itemize}

\item The number of clusters decreases with $\Gamma$ according to $N \propto 
\Gamma^{\approx -1.55 \pm 0.02}$ for the LCDMc scenario, though one should 
keep in mind that the different scenarios that are used do also have a 
different value of $\sigma_{8}$. 

\item The cluster peculiar velocity scales as $\Gamma^{\approx 0.32 \pm 0.01}$ 
for the LCDMc scenario and as $\Gamma^{\approx 0.38 \pm 0.01}$ for the LCDMb 
scenario. 
This is quite surprising as a larger $\Gamma$ indicates {\it less} 
large--scale power. 

\end{itemize}

\end{itemize}

\par In summary, we conclude that $\sigma_{8}$ has the largest influence on 
the cluster properties. 
This is not surprising because the mass within the virial radius of a rich 
cluster is very close to the mass enclosed within a sphere of $8 h^{-1}$ Mpc 
in the unperturbed Universe (e.g., Evrard 1989). 
Almost all cluster properties change if $\sigma_{8}$ is varied. 
$\Omega_{0}$ has a large impact on the cluster number density, mass, and 
peculiar velocity. 
In addition, for the number--selected cluster set, relatively more 
high--$\sigma_{\mbox{los}}$ clusters are expected if $\Omega_{0}$ is larger. 
More power on larger scales produces somewhat more elongated clusters and 
larger cluster peculiar velocities. 
The other parameters, the spectrum and $h$, correlate less strongly with the 
cluster properties. 

\par The cluster peculiar velocity is the cluster property that depends on the 
largest number of parameters of the fluctuation scenario. 
Two difficulties affect the determination of the most consistent cosmological 
scenario purely on the basis of cluster peculiar velocities. 
First, the large sensitivity of cluster peculiar velocities on all 
cosmological parameters make it very hard to disentangle these parameters 
and determine each of them separately. 
Fortunately, quite a few relations between cosmological parameters are known 
from other studies. 
E.g., Big Bang Nu\-cleo\-syn\-the\-sis and the fact that clusters cannot 
consist of more than 100\% baryons put severe limits on the Hubble parameter 
$h$ as a function of $\Omega_{0}$ (David, Jones \& Forman 1995). 
Secondly, cluster peculiar velocities are very hard to determine 
observationally. 
Mould {\it et al.} (1991, 1993), e.g., quoted errors of 300 to 800 km 
s$^{-1}$. 
Different studies sometimes yielded very different values for $v_{\mbox{pec}}$ 
for the same cluster. 
Very recently, Giovanelli {\it et al.} (1997) obtained the peculiar velocities 
for a sample of 22 groups and clusters. 
Although the sample is rather small, the uncertainties are considerably 
smaller (about 150 km s$^{-1}$) than those in previous studies and it will 
be very worthwhile to extend this data set to a larger number of clusters. 

\par As the cluster number density depends on the normalization $\sigma_{8}$, 
on the density parameter $\Omega_{0}$ and on the shape of the power spectrum, 
it can be used to discriminate between different scenarios (White {\it et al.} 
1993, Eke {\it et al.} 1996). 
However, the cluster number density in the simulations depends on the exact 
definition of a cluster. 
The most straightforward definition is, of course, to apply a mass threshold 
and consider all objects with a mass larger than this threshold to be 
clusters. 
But it is hard to get a reliable mass estimate of a cluster from galaxy 
positions and velocities. 
Better mass estimates may be obtained from X--ray measurements and 
gravitational lensing. 
Another way to define a cluster may be to use the cluster line--of--sight 
velocity dispersion and use only those objects which have a 
$\sigma_{\mbox{los}}$ larger than some well--chosen value. 
With the large redshift surveys coming up in the very near future, and with 
the ENACS--survey (Katgert {\it et al.} 1996) already being completed, this 
may prove to be a suitable manner to define a cluster. 
However, one does not necessarily pick out the most massive objects because of 
the scatter in the $\sigma_{\mbox{los}}-M$--relation (see Figure 
\ref{sigma_m}). 

\par It is promising to note that most of the cluster properties discussed 
do not depend critically on the cluster definition. 
That is, most results are qualitatively the same for the mass--limited and the 
number--selected cluster sample. 
However, when comparing to observations one should try to construct a model 
cluster catalogue that is complete in, e.g., richness or X--ray temperature. 

\par In Paper II, we will compare the properties of the 144 most massive 
clusters in the simulations with many observations. 
We then try mimic as closely as possible the way in which real clusters are 
observed. 
The results of the comparison with observations, together with the results of 
other studies, can then be used to find the cosmological scenario which is 
most consistent with all cluster properties. 
If necessary, the scaling relations can be used to 'interpolate' between 
scenarios. 
For the best--fitting scenario, a higher resolution cluster catalogue can then 
be constructed for which each cluster will be simulated individually at high 
resolution. 


\subsection*{Acknowledgments}

We are grateful to Edmund Bertschinger for providing us with his 
P$^{3}$M--code, to the National Computing Facility and SARA Amsterdam for 
giving us a large amount of computing time on the Cray C90. 
We also wish to thank Frank Robijn and SARA Amsterdam for providing us with 
extra computing time that made it possible to run an extra scenario. 
Furthermore, Peter Katgert and Tim de Zeeuw are acknowledged for reading 
earlier versions of this manuscript and for many useful suggestions. 
EvK acknowledges an European Community Research Fellowship as part of the 
HCM program.

\label{lastpage}

\end{document}